\documentclass{aa}  

\usepackage{graphicx}
\usepackage{txfonts}
\usepackage{hyperref}
\hypersetup{
    colorlinks=true,
    linkcolor=blue,
    filecolor=blue,      
    urlcolor=blue,
    citecolor=blue,
    pdftitle={Overleaf Example},
    pdfpagemode=FullScreen,
    }
\usepackage{xcolor}

\defcitealias{Spencer2018}{S18}
\defcitealias{Spencer2017}{S17}

\begin{document}

   \title{Binary star population of the Sculptor dwarf galaxy\thanks{ESO programme ID: 593.D-0309}}

   \author{
    José María Arroyo-Polonio \inst{1, 2}\thanks{E-mail: jmarroyo@iac.es (IAC)},
    Giuseppina Battaglia\inst{1, 2},
    Guillaume F. Thomas\inst{1, 2}, \\
    Michael J. Irwin\inst{3}, 
    Alan W. McConnachie\inst{4},
    \and Eline Tolstoy\inst{5}.
    \fnmsep
    }

   \institute{Instituto de Astrofísica de Canarias, Calle Vía Láctea s/n E-38206 La Laguna, Santa Cruz de Tenerife, España.    
         \and
             Universidad de La Laguna, Avda. Astrofísico Francisco Sánchez E-38205 La Laguna, Santa Cruz de Tenerife, España.
        \and
            Institute of Astronomy, Madingley Road, Cambridge CB3 0HA, UK.
        \and
            NRC Herzberg Astronomy and Astrophysics, 5071 West Saanich Road, Victoria, BC, V9E 2E7, Canada
        \and
            Kapteyn Astronomical Institute, University of Groningen, PO Box 800, 9700AV Groningen, the Netherlands.
             }

   \date{Accepted on July, 8th 2023}

 
  \abstract
   {}
   {We aim to compute the binary fraction of "classical" dwarf spheroidal galaxies (dSphs) that are satellites of the Milky Way (MW). This value can offer insights into the binary fraction in environments that are less dense and more metal-poor than our own galaxy. Additionally, knowledge of the binary fraction in dwarf galaxies is important with respect to avoiding overestimations of their dark matter content, inferred from stellar kinematics.}
   {We refined an existing method from the literature, placing
    an emphasis on providing robust uncertainties on the value of the binary fraction. We applied this modified method to a VLT/FLAMES dataset for Sculptor, specifically acquired for the purpose of velocity monitoring of individual stars, as well as to literature datasets for other six MW "classical" dSphs. In all cases, the  targeted stars were mainly red giant branch stars (RGBs), with expected masses of around 0.8 M$_{\odot}$. The VLT/FLAMES dataset offers the most precise binary fractions compared to literature datasets, due to its time baseline of 12 years, along with at least nine repeated observations for each star.}
   {We found that the binary fraction of Sculptor is 0.55$^{+0.17}_{-0.19}$. We find that it is important to take into account the Roche lobe overflow for constraining the period distribution of binary stars. In contrast to what has recently been proposed in the literature, our analysis indicates that there is no evidence to support varying the properties of the binary stellar population or their deviations from those established for the solar neighborhood, based on the sample of MW dSphs  analyzed here.}
   {}


   \authorrunning{Arroyo-Polonio J.-M. et al.}
   \titlerunning{The binary star population of Sculptor}
   \maketitle 
   
%

\section{Introduction}

The majority of Local Group (LG) dwarf galaxies (DGs) are pressure-supported \citep[e.g.,][]{wheeler2017} and devoid of HI gas \citep[e.g.,][]{putman2021}, which means that their dynamical mass is inferred from measurements of the velocity dispersion of their stellar component. While some first measurements of the velocity dispersion on the plane of the sky exist for an handful of DGs satellites of the Milky Way (MW) \citep[][]{massari2018, massari2020, delpino2022}, the vast majority of dynamical mass estimates of LG DGs rely on the velocity dispersion measured from line-of-sight (LoS) velocities, $\sigma_{los}$.  Since the first such study by \citet{Aaronson1983} for the MW satellite Draco, the LoS velocity dispersion has been measured for more than 80 LG DGs, with values ranging from $\sigma_{los}=0.9^{+0.6}_{-0.5}$ km~s$^{-1}$ for Tucana~III to $35 \pm5$ km~s$^{-1}$ for NGC~205 for  systems, where $\sigma_{los}$ could be resolved \citep[see recent review by][and references therein]{battaglia2022nature}. A result that has been consolidated over the years states, contrary to globular clusters, the $\sigma_{los}$ of LG DGs is significantly higher than the values that one expects from the contribution of baryons, with dynamical mass-to-light ratios $(M_{1/2}/L_{1/2})_{\rm dyn} \gtrsim$ 10 in solar units\footnote{The luminosity quoted refers to the V- band.} within the 3D half-light radius for the vast majority of these galaxies. In addition, the dynamical mass-to-light-ratio is inversely proportional to the DG luminosity \citep[e.g.,][]{MMateo1998, mcconnachie2012, Walker2012, Strigari2013, battaglia2022nature}, with $(M_{1/2}/L_{1/2})_{\rm dyn} \gtrsim$ 1000 for the so-called ultra-faint dwarfs (UFDs)\footnote{Here, we follow \citet{Simon2019} and define as UFDs or classical dwarf galaxy those DGs fainter and brighter than M$_V = -7.7$, respectively. The DGs devoid of gas are named dwarf spheroidals (dSphs).}. In standard Newtonian dynamics, these high mass-to-light-ratios are explained by the fact that dwarf galaxies are embedded in dark matter halos several order of magnitudes more massive than the luminous component. It is still possible (particularly for the faintest systems) that a non-negligible fraction of the velocity dispersion measured in DGs is a consequence of other effects, one of them being the presence of unresolved binary stars \citep[e.g., see early studies by][]{Hargreaves1996,Olszewski1996}. 

It has been shown that the effect of unresolved binary stars is minimal for galaxies with high intrinsic velocity dispersions \citep{Hargreaves1996,Quinn}, namely, in the regime of "classical" DGs. For example, a binary fraction equal to 1 (under the assumption of period, mass and eccentricity distributions as in the solar neighborhood) would inflate an intrinsic LoS velocity dispersion $=8$ km s$^{-1}$ to 9 km s$^{-1}$ \citep{Spencer2017}. However for galaxies with low intrinsic dispersions, $\lesssim2$ km~s$^{-1}$, the same study shows that the measured $\sigma_{los}$ can be 1.5–4 times higher than the intrinsic dispersion already with a modest binary fraction of 0.3. 

The value for which the measured $\sigma_{los}$, and, hence, the inferred dynamical mass-to-light ratio, inflated by binary stars, is strongly dependent on the fraction of stars located in binary systems, as has also been concluded by \citet{pianta2022}.\footnote{The differences in the results between this study and others that assess the impact of non-resolved binaries on 
the $\sigma_{los}$. \citep{mcconachie2010, Spencer2018} arise from the difference in the period distribution they use.} In particular, if the fraction of binaries with periods lower than 10 years exceeds 5\%, there is a non-negligible ($>$5\%) chance that the values of $\sigma_{los}$ from single-epoch observations might be completely due to binary systems for, for instance, Segue~1, Segue~2, Willman~1, Bootes~II, Leo~IV, Leo~V, and Hercules \citep{mcconachie2010}. Thus far, in  studies of UFDs for which multi-epoch spectroscopic measurements have been acquired, removing (or accounting for) stars showing obvious velocity variations reduces the value of the measured $\sigma_{los}$ \citep[e.g.,][]{Martinez2011, Ji2016, Kirby2017, Minor2019, Buttry2022}, even though it is not enough to bring the dynamical mass-to-light ratio in line with expectations from the baryonic component only. Nonetheless, it remains of interest to determine how strongly the inferred dynamical mass-to-light ratios of individual UFDs might be inflated by unresolved binary stars.

Unlike the Solar neighborhood, the properties and the fraction of binary stars is generally unknown in LG dwarf galaxies. In the Milky Way, \citet{Duquennoy1991} found an overall binary fraction of $\sim 2/3$ for stars near the middle of the main sequence, with the binary fraction decreasing from $f\simeq 0.5-0.6$ for very metal-poor stars down to $f \simeq 0.1$ for the stars with super-solar metallicities \citep{Raghavan2010,Badenes2018,Moe2019}. Moreover, \citet{Hettinger2015} found that disk stars, mainly metal-rich, are 30\% more likely to belong to a short-period ($<12$ days) binary system than metal-poor halo stars.

Regarding dwarf galaxies, the binary fraction has been investigated in a few of the Milky Way satellite galaxies, mostly in the so-called ''classicals'' dSphs \citep[e.g.,][]{Quinn, Spencer2017, Spencer2018}, but also in some UFD \citep[e.g.,][]{Martinez2011, Minor2019}. Typically, these studies have to make the assumption that properties of the population of binary stars in DGs, such as their period and mass ratio distributions, are the same as in the solar neighborhood. What makes it particularly challenging is that this type of explorations necessitate accurate and precise LoS velocity determinations from multi-epoch observations of a large number of stars, as pointed out by \citet{Martinez2011}, for instance. \citet{Quinn} found (with the notable exception of Carina) that three ''classical'' MW dwarf galaxies (Fornax, Sculptor, and Sextans) have a uniform binary fraction of $f\simeq 0.5$, similar to the value found for the metal-poor stars in the Milky Way. However, these results were recently contested by \citet{Spencer2018} (hereafter \citetalias{Spencer2018}), who revisited the binary fraction in these galaxies plus three others (Ursa Minor, Draco and Leo II). They found that these seven dwarf galaxies cover a wide range of binary fractions; they rejected at a $99\%$ level the universality of the binary fraction in dwarf galaxies (i.e., that they have the same binary fraction within a range of width 0.2), provided that they share similar period distributions; or conversely, that if they share similar binary fractions, then their period distributions need to be significantly different. \cite{Bonidie2022} examined the velocity variation of stars across different epochs in the Sagittarius DG and compared them with a sample of stars from the MW of similar stellar atmospheric parameters. These authors found that Sagittarius stars display larger velocity variations than the comparison Milky Way stars, so they ruled out all the possible explanations -- leaving just the notion of a larger binary fraction for close binaries.

If the results by \citet{Spencer2018} were confirmed, this would exclude the possibility of using the binary stellar population of dSphs as templates for those in UFDs; however, constraining its properties is even more challenging due to the scarcity of target stars bright enough to be followed up with current facilities. 
In this paper, we  revisit the determination of the binary fraction in these galaxies. As aptly highlighted in \citet{Quinn}, knowing the properties of binary stars in MW DGs is also important in constraining star formation theories in galactic environments of low metallicity and covering a variety of star formation histories.
Moreover, since DGs are more diffuse than globular clusters, they present a lower chance of stellar encounters, making DGs a more controlled environment in which to test stellar formation theories, especially since formation models predict a binary fraction ranging from $f=0.05$ \citep{Jarrod2007} to $f=1.0$ \citep{Ivanova2005} for them.

In this work, we first focus on the Sculptor dwarf spheroidal (dSph), presenting results from a homogeneous spectroscopic dataset for 96 red giant branch (RGB) stars obtained at the Very Large Telescope (VLT) with the multi-object spectrograph Fiber Large Array Multi Element Spectrograph (FLAMES). This dataset was based on the same instrument set-up between 2003 and 2015, with 9 to 12 epochs per star. Given the results obtained for this galaxy, we propose some modifications to Bayesian methods existing in the literature \citep{Spencer2018}, so as to obtain more robust uncertainties. Using this new method, we re-analize the binary fraction of the six other DGs studied in \citetalias{Spencer2018} (Draco, Ursa Minor, Sextans, Carina, Leo II, and Fornax).

The paper is organized as follows. In Sect.~\ref{sec:Data}, we introduce the datasets used in this work, including the new multi-epoch VLT/FLAMES dataset for Sculptor. In Sect.~\ref{sec:methodology}, we describe the methodology employed and present the changes with respect to \citetalias{Spencer2018}. Our results are presented in Sec.~\ref{results}, where we first determine the binary fraction in Sculptor (Sect.~\ref{sec:scl}) and then those of the other six MW dSphs (Sect.~\ref{sec:lit}).  In Sect.~\ref{sec:outlook}, we present an outlook for what future observations with next-generations multi-object spectrographs such as WEAVE will contribute to studies of the population of binary stars in MW dSphs \citep{dalton2012, Jin2023}. Finally, we draw our conclusions in Sect.~\ref{sec:conclusions}. In Appendix~\ref{sec:appendix1}, we discuss whether there might be a common binary fraction that would cover all the DGs.

\section{Data} \label{sec:Data}
Here, we  present a new set of FLAMES/GIRAFFE observations for the Sculptor dSph (Sect.~\ref{sec:data_scl}), followed by the literature datasets used for Draco, Ursa Minor, Leo~II, Carina, Fornax, and Sextans (hereafter, Dra, UMi, Leo~II, Car, Fnx, and Sxt), as explained in Sect.~\ref{sec:data_lit}. 

\subsection{FLAMES/GIRAFFE Sculptor dataset}\label{sec:data_scl}
Our analysis of the binary star fraction in the Sculptor dSph relies on LoS velocities of individual RGB stars measured at different times over a large time baseline. The LoS velocities are obtained from spectroscopic data acquired at the VLT with the instrument FLAMES in GIRAFFE mode. FLAMES is perfectly suited for the objective of this work as it is known to be a very stable spectrograph \citep{FLAMESPerformance}. The section is organized as follows. In Sect.~\ref{sec:maincha}, we describe the main characteristics of the data and briefly comment on the data-reduction process and the determination of LoS velocities. In Sect.~\ref{sec:zeropoint}, we perform an analysis of the possible zero-point velocity offsets between the different observations of the same sources. In Sect.~\ref{sec:robustness}, we analyze the robustness of the velocity uncertainties.

\subsubsection{Observations, data-reduction, and determination of velocities} \label{sec:maincha}
\begin{figure*}
    \centering
        \includegraphics[width=1.65\columnwidth]{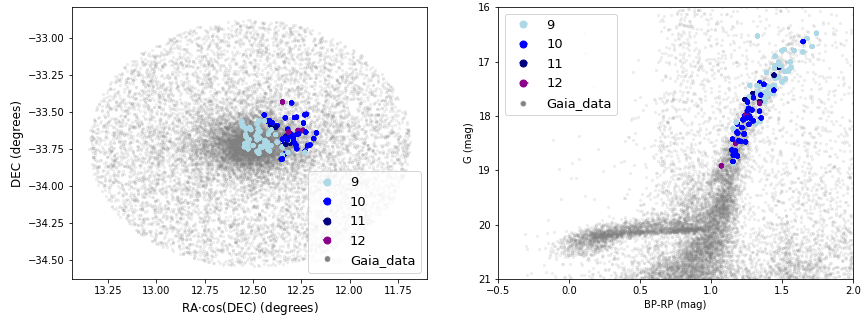}
        \caption{Distribution of Gaia eDR3 sources (in gray) within a 50 arcmin radius around the central coordinates of the Sculptor dSph, with overlaid the 96 targets of the FLAMES-GIRAFFE multi-epoch observations (large dots). Left panel: Distribution  on the sky. Right panel:\ Distribution on the color-magnitude diagram. The color indicates the number of repeated observations (see legend).}
        \label{Sculptorsky}
    \end{figure*}

Our targets are 96 stars with {\it Gaia} early Data Release 3 (eDR3) \citep{Gaia2,Gaia1} G-magnitudes, approximately between 16.5 and 19 and located within 25 arcmin from Sculptor center (see Fig. \ref{Sculptorsky}). These were selected among the likely Sculptor members from the datasets presented in \cite{tolstoy2004, Battaglia2008,Starkenburg2010}\footnote{All of them have a probability membership above 85\% in the catalog by \citet{battaglia2022}} and followed up on as part of an ESO monitoring program over two years (ID: 593.D-0309; PI: Battaglia). Specifically, the monitoring consisted of eight repeated observations of 2700 seconds, between 2014 and 2015, with the same instrument set-up and fiber configuration, in order to reduce the source of errors on the repeated measurements of the LoS velocities. The grating used was LR8, covering the region of the near-infrared Ca~II Triplet (CaT). The choice of placing the pointing at a location displaced from Sculptor center was due to the wish of probing a region with a mean metallicity more akin to that of ultra-faint dwarf galaxies. In the monitoring programme, we aimed for observations spaced approximately logarithmically over time. For the actual observation dates, see Table~\ref{Tab:offsets}.

\begin{figure}
    \begin{center}
    \centering
        \includegraphics[width=0.85\columnwidth]{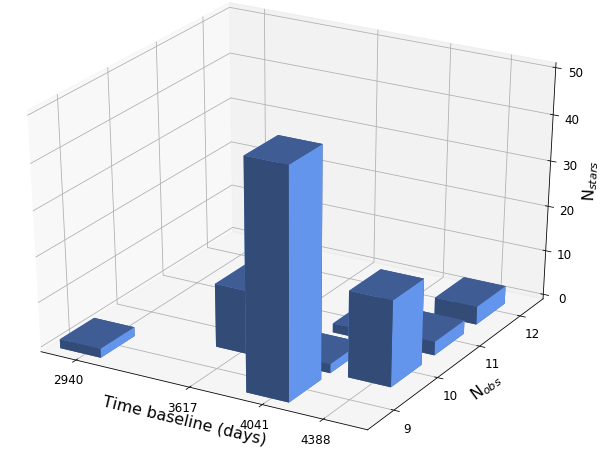}
        \end{center}
        \caption{Distribution of time baseline and number of repeated observations for the 96 stars that are part of our Sculptor dataset.}
        \label{fig:timebs}
    \end{figure}

In the earlier programs  the targets were drawn from, the spectra for these 96 stars were acquired over multiple pointings, between 2003 and 2007 -- again, with the same instrument and grating, but with a different fiber configuration. In these nights not all the same 96 stars were observed, but each of them has at least one extra observation in the 2003-2007 period. Therefore, the sample consists of 96 stars, with at least nine repeated measurements, with a minimum time baseline of about eight years (2940 days). Figure~\ref{fig:timebs} summarizes the number of stars for a given number of repeated measurements and time-baseline. Table~\ref{Tab:offsets} lists the dates of the observation and the number of stars proceeding from those exposures. 

The approach adopted for the data reduction, extraction of the 1D spectra, sky-subtraction, and determination of the velocities is presented in \cite{Tolstoy2023},  to which we refer for more details. Of particular relevance here is the set of LoS velocities and uncertainties adopted. In Tolstoy et al. two methods for determining LoS velocities were used: 1) cross-correlation with a template with zero continuum and three Gaussians mimicking the CaT lines in absorption, as in \citet{Battaglia2008}, with the uncertainties being derived from the scatter of the velocities returned by each of the 3 Gaussians; 2) a maximum-likelihood-method where 3 parameters ($v_{los}$, $\Delta v_{los}$ and best fit Gaussian width for blurring  the delta function of the CaT template) are explored to obtain the most likely combination. In order to decide which set to adopt, we did our own error analysis for the 96 stars of the monitoring program, since it is a key parameter in our work. Here, we adopted the cross-correlation set, since it has more reliable errors.

As a confirmation that the targets of the monitoring program are members to Sculptor, we checked that their average LoS velocity ranges between 89 km s$^{-1}$ and 137 km s$^{-1}$, which is well within 3-$\sigma$  from the systemic LoS velocity of Sculptor \citep[the systemic velocity being v$_{\rm los, sys}=110 \pm 0.5 $ km s$^{-1}$ and the LoS velocity dispersion $\sigma_{\rm los, sys}=10.1\pm0.3$ km s$^{-1}$, see][]{Battaglia2008}. These values are in agreement with what is found in \cite{Tolstoy2023} with an updated version of this dataset that features a total of 1604 probable member stars with velocity measurements, v$_{\rm los, sys}=111.20 \pm 0.25 $ km s$^{-1}$.

\begin{figure}
    \begin{center}
    \centering
        \includegraphics[width=0.95\columnwidth]{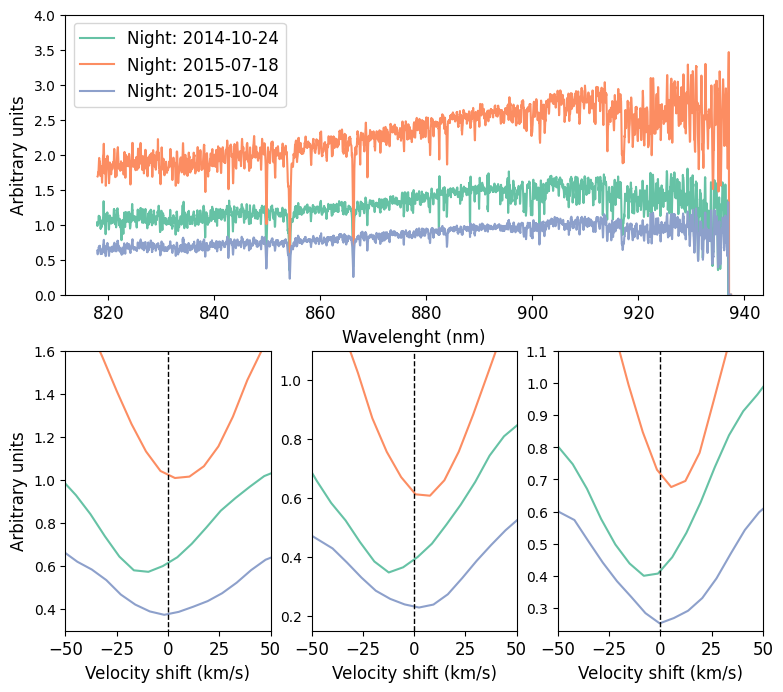}
        \end{center}
        \caption{Spectra of an example Sculptor member star, observed during three different nights, as indicated in the legend. The top panel shows the whole spectra. The bottom panels show a zoom-in view of the CaT lines, with the x-axis indicating the velocity shift with respect to the rest wavelength (indicated with dashed lines) of a given CaT line.}
        \label{fig:espec}
    \end{figure}

The monitoring program was designed to deliver spectra with intermediate-to-high signal-to-noise ratio (S/N) and, therefore, precise LoS velocities. The minimum S/N in the spectra observed between 2014 and 2015 is around 20; for the spectra from the previous nights, 2003-2007, the minimum S/N is around 5.6. These spectra produce a mean velocity error of $\sim$ 1  km s$^{-1}$. In the final dataset, namely, taking into account all observations, the individual spectra display a mean S/N of 56. In Fig.~\ref{fig:espec}, we show three spectra of the same star at different dates, in which we can see velocity variation.


\subsubsection{Correction for zero-point offsets} \label{sec:zeropoint}
Even though the observations were carried out with the same instrument and set-up, systematic differences in the $v_{\rm los}$ between the exposures taken during different nights can still exist. Those systematic errors would increase the differences between the $v_{los}$ for a given star, potentially inducing an inflated inferred binary fraction. 

\begin{figure}
        \includegraphics[width=\columnwidth]{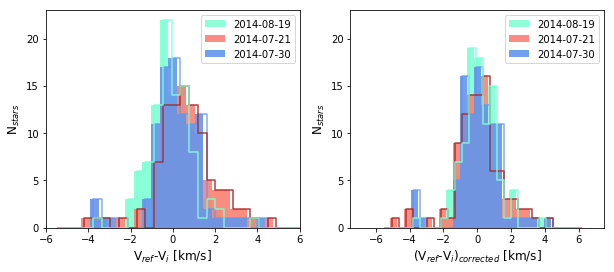}
        \includegraphics[width=\columnwidth]{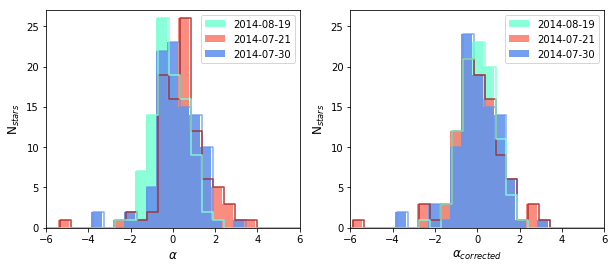}
        \caption{Distributions of velocity differences and $\alpha$. Upper panels: Histograms of the difference between velocities for the different nights with respect to the reference night, as indicated in the legend, before the zero-point offset correction (left) and after the correction (right). Lower panels: Same as the upper panels but for $\alpha$ instead of the velocity difference. Only three nights have been shown for ease of visualization, but all nights have been corrected.}
        \label{Correccionoffsets}
    \end{figure}

Here, we compare the LoS velocities of each star $j$ obtained in each exposure $i$, $v_{i,j}$, with the LoS velocities measured for the same star during a reference night, $v_{ref,j}$. For simplicity, hereafter, we drop the subscript $j$. As reference night, we adopt that with the best S/N and G-magnitude relation, which was on 2014-08-19.
An example of the distribution of the differences, $v_{ref}-v_{i}$, is shown in the upper left part of Fig. \ref{Correccionoffsets}, where we present the case only for three nights in order to better visualize the results. 
The typical zero-point shifts, measured as the mean values of $v_{ref}-v_{i}$ (listed in the second column of Table \ref{Tab:offsets}), are typically rather small, around 0.18 km s$^{-1}$ for the nights of the monitoring program; nonetheless, we subtracted those shifts from the velocities of the stars observed in a given night. Examples of the distributions corrected for these offsets are given in the upper right panel of Figure~\ref{Correccionoffsets}: now all the distributions are centered on 0 km s$^{-1}$. Since we do not have enough statistics for the 2003-09-30 night (only 1 out of our 96 stars was observed that night), those velocities have not been corrected. The other case with low statistics is the night of 2007-09-16. For this one, we decided to correct by offsets. However, this is not a problem since we checked that removing any of those nights, does not impact the results at all.

\subsubsection{Robustness of the velocity uncertainties} \label{sec:robustness}
In order to check whether the uncertainties on the $v_{los}$ measurements are robust, we examined the distribution of the differences between the $v_{los}$ of each star in a given night and the reference night, $v_{ref}-v_{i}$, weighted by the respective velocity uncertainties:    
    \begin{equation}
        \alpha=\frac{v_{ref}-v_{i}}{\sqrt{\left(\Delta v_{ref}\right)^{2}+\left(\Delta v_{i}\right)^{2}}}
    ,\end{equation}
   
\noindent where $\Delta$ v$_{ref}$ and $\Delta$ v$_{i}$ are the velocity uncertainties of a given star $j$ in the nights $ref$ and $i$. Table~\ref{Tab:offsets} shows the scaled median absolute deviation ($sMAD$)\footnote{defined as $sMAD=1.48 \cdot$ median$|\alpha_{i}-$median$(\alpha_{i})|$} for the parameter $\alpha$ for each given night, after applying the offsets' correction (col. 4). We prefer the $sMAD$ to the standard deviation as the former is more robust to the presence of outliers. In the lower part of Fig. \ref{Correccionoffsets}, we can see examples of the distribution of $\alpha$ before and after the correction. For a set of measurements where the only source of errors are statistical ones, we expect the distribution of $\alpha$ to be a Gaussian centered on 0 and with $sMAD$ of 1; values of $sMAD$ lower or larger than 1 would indicate over- or under-estimated velocity uncertainties, respectively. 
In Table \ref{Tab:offsets}, we may notice that the $sMAD$  of the distribution of $\alpha$ (Col. 4) for the nights between 2014 and 2015, where the same exact fiber configuration was used, is systematically lower than 1, which would point to the uncertainties being overestimated. Dividing them by a factor 1.15 leads to sMAD values centered around 1 (Table~\ref{Tab:offsets}, Col. 5), so we adopted this correction factor for our uncertainties. However, after the correction, we still obtain a range of $sMAD$ between 0.81 and 1.23. There are a couple of reasons as to why the $sMAD$ might differ from 1 even after correcting the velocity uncertainties: a) in this case, we are treating the dataset as if there were no binary stars and b) we are studying a sample of 96 points as a statistical distribution.

    Focusing on point b), we sampled 20000 Gaussian distributions centered on 0 and with standard deviation 1 with 96 points and then computed their $sMAD$. The final distribution of the values of $sMAD$ is a Gaussian 
    centered on 1 and with a width of $\sigma$ = 0.12. Therefore, all the values of sMAD we obtained after correcting the velocity uncertainties are within 2$\sigma$ of the $sMAD$ expected for a distribution sampled with 96 points. The results from the 2003-2007 nights have larger deviations from unity; however, in those cases, the number of stars in the exposure is lower than 96, hence it is natural to observe larger deviations\footnote{Here we neglect the case for the night of 2007-09-16, due to the small number statistics (5 stars).}. 
    Therefore, in general, our results are in line with the expectations under the conditions of the observations and we consider our velocity errors to be estimated as well, once they are corrected by a factor of 1.15. 

    The jitter due to the movement of material in the surface of the RGB stars is also a factor to take into account when it comes to uncertainties. This is because it can occur on short timescales, on the order of minutes \citep{Yu2018} and, therefore, we are going to measure it as it was noise in the velocity measurements. The effect of jitter on $v_{los}$ is a function of the stars' surface gravity \citep[e.g.,][]{Hekker2008}. Therefore, we have cross-matched our dataset with APOGEE to obtain a subsample of stars with known values of surface gravity. The stars in our dataset have $2 > $ $\log$(g) [cgs] $ > 0.2$, which correspond to jitter velocities between $0.12 < $ v$_{jitter}$[km s$^{-1}$] $< 1.40 $. On average, the jittering velocities of our sample are around 0.5 km s$^{-1}$, so it should already be accounted for in our errors, and in the correction factor that we have applied.

    We also checked for possible trends in the velocity determination, that is, those produced by the wavelength calibration of the spectra, by searching for trends in $\alpha$ as a function of v$_{ref}$. In cols. 6 and 7 from Table \ref{Tab:offsets}, we can see the slope $m$ of the best linear fit of $\alpha$ as a function of v$_{ref}$ and the dispersion coefficient, $r^2$. We find that both $m$ and $r^2$ are close to 0 in all the cases, indicating that there is no increasing or decreasing trend of $\alpha$ with v$_{ref}$ and no relation between both variables. 
    
    \begin{table*}
\caption{Statistical properties of our dataset, divided in nights.}
\label{Tab:offsets}
\centering
\begin{tabular}{llllllllll}
\hline
Date & Julian date  & Mean [km s$^{-1}$]    & $sMAD$($\alpha$) & $sMAD$$_{corr}$($\alpha$) & m      & r$^2$  & $n_{obs}$ & Mean SNR \\
\hline

2003-09-29 & 2452911.54 & $-$0.13  &  1.22   & 1.38  & \hspace{2.7pt} 0.02   & 0.020  & 50     &  29.0\\
2003-09-30 & 2452912.54 & \hspace{2.7pt} 0.07       &  0      & 0     & -      & -      & 1    &  12.2\\
2004-09-10 & 2452892.84 &  \hspace{2.7pt} 0.90    &  1.42   & 1.19  & $-$0.05  & 0.020 & 56  &  51.8\\
2005-11-08 & 2453682.71 & $-$1.60   &  1.94   & 1.60  & $-$0.09  & 0.110  & 37   &  23.5\\
2007-09-16 & 2454359.67 & $-$6.07   &  2.10   & 0.40   & \hspace{2.7pt} 0.17   & 0.045  & 5    &  24.1\\
2014-07-20 & 2456858.86 & $-$0.03   &  0.83   & 0.99  & $-$0.02  & 0.012  & 96  &  53.7\\
2014-07-21 & 2456859.84 & \hspace{2.7pt} 0.41    &  0.80   & 0.92  & $-$0.01  & 0.001  & 96    &  53.3\\
2014-07-30 & 2456868.70 & $-$0.12   &  0.84   & 0.96  & $-$0.02  & 0.009  & 96  &  65.4\\
2014-08-16 & 2456885.71 & \hspace{2.7pt} 0          &  0      & 0     & \hspace{2.7pt} 0      & -      & 96 &  63.5\\
2014-08-19 & 2456888.71 & $-$0.21   &  0.71   & 0.81  & $-$0.01  & 0.003  & 96  &  61.3\\
2014-10-24 & 2456954.59 & \hspace{2.7pt} 0.10    &  0.95   & 1.10  & $-$0.02  & 0.008  & 96    &  53.6\\
2015-07-18 & 2457221.86 &\hspace{2.7pt} 0.33    &  0.90   & 0.92  & $-$0.03  & 0.017  & 96    &  64.2\\
2015-10-04 & 2457299.67 & \hspace{2.7pt} 0.23    &  1.13   & 1.23  & $-$0.01  & 0.001  & 96    &  61.0\\
\hline
\end{tabular}
\tablefoot{Col.~1 lists the night of observation. Col.~2 the mean of the difference of velocities between that night and the reference night (2014-08-16). Col.~3 the standard deviation of the difference between velocities before the offsets' correction. Col.~4 the sMAD for the distribution of the parameter $\alpha$ before the correction of the velocity uncertainties. Col.~5 the sMAD for the distribution of $\alpha$ after the offset correction. Col.~6 the slope of the best linear fit between the difference between velocity and the velocity in the reference night used to compute it. Col.~7 lists the dispersion coefficients of the data in the same velocity space; for the 2014-2015 observations, Col.~ 8 lists the number of stars in each exposure, while for the 2003-2007 observations, it gives the number of stars overlapping to those observed in the monitoring program and proceeding from a given night. Finally, Col.~9 lists the mean S/N of the spectra at each night.}

\end{table*}

\begin{table}
        \centering
        \caption{Parameters of the galaxies considered in this study.}
        \label{tab:Paramet}
        \begin{tabular}{lccc|ccc} 
                \hline
                  & $L_{c}$  & $\sigma_{los}$ & $\lambda$ & t  & $a_{max}$\\
                  & $\left[\text{L}_{\odot}/\text{pc}^{3}\right]$ & $\left[\text{km s}^{-1}\right]$ & $\left[\text{pc}^{-3}\right]$ & [Gyr] & [AU]\\
                \hline
                Sculptor & 0.009 & 10.1 & 0.36  & 12 & 551 \\
                Draco & 0.008 & 9.0 & 0.32  & 10 & 678 \\
                Ursa Minor & 0.006 & 8.0  & 0.24  & 13 & 728 \\
                Leo II & 0.029 & 7.4 &  1.16 & 9 & 414 \\
                Carina & 0.006 & 6.6 & 0.24  & 8 & 1022 \\
                Fornax & 0.018 & 11.4 & 0.72  & 8 & 449 \\
                Sextans & 0.02 & 8.4 & 0.80  & 13 & 389 \\
                \hline
        \end{tabular}
 \tablefoot{ Col.~1: Galaxy name. Col.~2: Central luminosity from \protect\cite{MMateo1998}. Col.~3:  LoS velocity dispersion of the stellar component as listed in \protect\cite{battaglia2022nature}. Col.~4: Average number density of stars. Col.~5: Typical age of the stellar population from \protect\cite{Bettinelli2019, Aparicio2001, Carrera2002, DeBoer2014, Betinelli2018, Deboer2012, Mighell1996}, respectively. Col.~6: Mean-free path of a star in the galaxy (maximum value of the semi-major axis for a binary system).}
\end{table}

\subsection{Data from the literature}\label{sec:data_lit}

In this work, we also re-analyze spectroscopic datasets used in the literature for studies of the binary stellar population of MW classical dSphs. Specifically, we  use the datasets from Draco and Ursa Minor presented in \citetalias{Spencer2018}; the one from Leo~II presented in \citetalias{Spencer2017}; and the ones from Sextans, Carina, and Fornax presented in \cite{walker2009}. All of these were used by \cite{Spencer2018}, while \cite{Quinn} adopted only the datasets from \cite{walker2009}, available at the time.

\section{Methodology} \label{sec:methodology}
The methodology, we adopt for determining the binary fraction, $f,$ of the dwarf galaxy stellar population  largely follows that of \citetalias{Spencer2018}, but with three main differences, as highlighted in the next sections. Broadly speaking, it consists of producing mock datasets with the same characteristics as the observational data, in terms of number of stars, time sampling, and 
velocity uncertainties, for different values of $f$, and then comparing them with the observational data.

In Sect.~\ref{sec:Simulation}, we give the details of the simulation strategy, in Sect.~\ref{sec:Parameters}, we recall the formula for the $v_{los}$, of a component of a binary system and the parameters it depends on, and we discuss the distributions we use for these parameters. In Sect.~\ref{sec:MAP} we present the maximum a posteriori method (MAP) we follow to compare the simulated and the observational data. Finally, in Sect.~\ref{sec:changes}, we highlight the differences with respect to similar methodologies used for this same objective by \citetalias{Spencer2017} and \citetalias{Spencer2018}.

\subsection{Simulation strategy} \label{sec:Simulation}
The following steps to produce the simulated datasets are the same as published in \citetalias{Spencer2018}. We summarize them below.

(1) First, we chose a value for the binary fraction, f.

(2) For each observed star, we randomly labeled it as a binary or single star, according to the binary fraction in Eq. 1. In practice, this is done by randomly extracting a value from a uniform distribution between 0 and 1. If the number extracted is lower than f, the star is labelled as binary; otherwise, it is not. If it labeled as binary,  we randomly assign to it a set of the seven parameters extracted from the distributions in Sec.~\ref{sec:Parameters}.

(3) We computed the v$_{los}$ of the star at the time of each observation. If the star is a binary, we use Eq.~(\ref{eq:vlos_bin}), taking into account the time variation of the true anomaly along all the observation epochs. If the star is a single star, we assign it a velocity of 0 km s$^{-1}$. Here we neglect the change in $v_{los}$, due to the intrinsic motion of stars within the galaxy, as it is negligible on our time baseline.

(4) Next, we computed the Gaussian deviations of the velocities with a standard deviation equal to the observational error of each observation. We added this value to the one obtained in the third step for both binaries and single stars. We would like to highlight that it is not necessary to add the systemic motion of the galaxy, since we are interested in the temporal velocity variations of each given star in the sample with respect to its velocity at a reference time.

(5) We repeated steps three and four for all the stars in the dataset.

(6) Then, for statistics, we repeated steps two to five 10,000 times. At this point, we have 10,000 different simulated mocks for each binary fraction f.

(7) Finally, we repeated steps one to six for all the binary fractions we are interested in. We sample binary fractions between 0 and 1, with a step of 0.01.
As in \citetalias{Spencer2018}, the parameter that serves as the basis for this study is: 
\begin{equation}
    \beta=\frac{\left|v_{i}-v_{j}\right|}{\sqrt{\left(\Delta v_{i}\right)^{2}+\left(\Delta v_{j}\right)^{2}}},
\end{equation}

\noindent where the subscripts $i$ and $j$ denote different observations of the same star, $v$ is the measured velocity, and $\Delta v$ the error of this measurement. The value of this parameter represents how large the velocity variation is between the different observations of the same star, weighted by the velocity uncertainties. We compute this parameter for both the observed and simulated data ($\beta_{obs}$ and $\beta_{sim}$, respectively).  For a star with $n$ observations, there will be $\frac{n(n-1)}{2}$ different values of $\beta$. In our particular case, we have 96 stars observed between 9 and 12 times. This leads us to 3986 different values of $\beta$. In previous works, the observations produced 2015 $\beta$ for Draco \citep{Spencer2018}, 1278 for Ursa Minor \citep{Spencer2018}, and 723 for Leo II \citep{Spencer2017}, far less than with our observations. So, we can see that we have not only significantly more data, but also the uncertainties in our velocity measurements are better, as we saw in Sect. 2.1.

\subsection{Parameters distributions} \label{sec:Parameters}

Assuming that a binary system behaves as a two point mass distribution with only gravitational interaction, $v_{los}$, of one of the components, with respect to the system's center of mass, can be derived analytically \citep[e.g., see][]{GreenSphericalDynamics}: 
    \begin{equation} \label{eq:vlos_bin}
    v_{los}=\frac{q\text{ }\sin(i)}{\sqrt{1-e^{2}}}\left(\frac{2\pi Gm_{1}}{P(1+q)^{2}}\right)^{1/3}\left(\cos\left(\theta+\omega\right)+e\text{ }\cos(\omega)\right).
    \end{equation}
The equation contains four intrinsic ($q$, $m_1$, $P,$ and $e$) and three extrinsic ($i$, $\omega,$ and $\theta$) parameters. Among the intrinsic ones: $m_1$ is the mass of the primary star; $q=\frac{m_{2}}{m_{1}}$ is the ratio between the mass of the secondary star, $m_2$, and of the primary star $m_1$; $e$, the eccentricity of the orbit around the center of mass; and $P$ the period of the orbit. Of the extrinsic parameters: $\omega$ is the argument of the periastron, the angle corresponding to the point of the orbit that is closest to the center of mass; $i$ is the inclination, that is, the angle between the plane determined by the binary system's orbit and the plane perpendicular to the LoS; and $\theta$ is the true anomaly, the only parameter that changes with time, as it indicates the position of the star along the orbit.

\begin{figure}
        \includegraphics[width=\columnwidth]{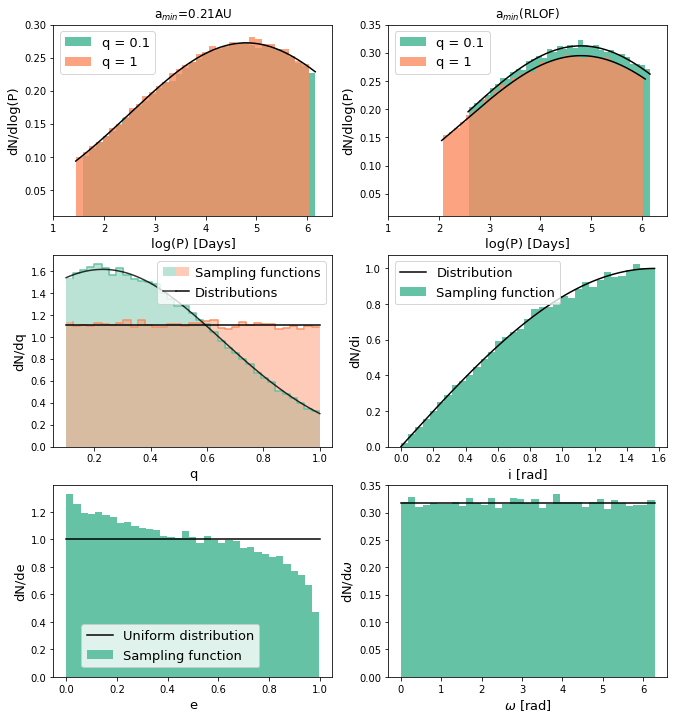}
        \caption{Probability distribution functions for the parameters defining the properties of the binary star populations. The black lines show the theoretical distributions while the colored histograms show the results of our sampling. Upper panels: Period distribution for the two limits in mass-ratios, as indicated by the legend, assuming a constant minimum orbital distance $a_{min}=0.21$ AU (left) or the taking into account the RLOF when computing $a_{min}$ (right). The limits for each mass-ratio are plotted with vertical lines. Middle left panel: Distributions adopted for the mass ratio (green: GS distribution from \protect\cite{Duquennoy1991}; salmon: uniform distribution). Middle right panel: Distribution of inclinations. Lower left panel: Eccentricity distribution; the differences between the uniform theoretical distribution and the one sampled are due to the dependence of $e$ on the semi-minor axis and therefore, the orbital period, as explained in the main text. Lower right panel: Distribution of the argument of the periastron.}
        \label{fig:Distribuciones}
    \end{figure}

We adopted the same distributions used in previous studies of binary fractions of DGs \citep{Spencer2017,Spencer2018,Quinn}, so that we can directly compare our results with theirs. We adopted the following distributions (see also Fig. \ref{fig:Distribuciones} 1). For the mass of the primary star,
$m_{1}$,  we  used a fixed value of m$_{1}$ = 0.8 M$_{\odot}$, since this is the typical mass expected for the majority of our target old and metal-poor RGB stars \citep[see e.g.,][]{Martig2016}.\footnote{Small deviations around this value of the order of 0.1 M$_{\odot}$ do not change the results.} For the mass measurement, $q$,  we  used a Gaussian distribution centered on $\mu_q =0.23$ and with dispersion $\sigma_{q}$ = 0.42 (see upper left panel of  Fig.~\ref{fig:Distribuciones}), as found by \citet{Kroupa1990}. It has been shown to fit correctly the distribution of mass ratios for G-dwarfs belonging to binary systems in the solar neighborhood \citep{Duquennoy1991}. However, we also tested a uniform distribution, as \citet{Kroupa1990} also showed that the distribution of mass ratios is flattened for binaries with short periods. This is assuming the uniform distribution yields very similar results to the Gaussian one. As for $P$, most studies about binary systems in the solar neighbourhood have shown that the observed period distribution is a log-normal function \citep{Duquennoy1991,DistP35,DistP58}:
    \begin{equation}
     f(\log P [d])\varpropto exp\left(\frac{-\left(\log P [d]-\overline{\log P [d] }\right)^{2}}{2\sigma_{\log P [d]}^{2}}\right),   
    \end{equation}
    where $\overline{logP}$ is the mean of the distribution, with the period in units of days, and $\sigma_{logP}$ is its width. 
    
    Even though previous works \citep{Quinn,Spencer2017,Spencer2018} have shown that the choice of the mean of the distribution and $\sigma_{logP}$ is crucial for the determination of the binary fraction in dwarf galaxies, the available datasets do not yet allow for such parameters in dwarf galaxies to be inferred, and strong degeneracies are present between the values of these parameters and the binary fraction that best reproduces the observational data. Throughout the article, we adopt the typical baseline case of $\sigma_{logP} = 2.3$ and $\overline{\log P\left[\text{d}\right]}=4.8\text{ }\left(\approx180\ \text{years}\right)$ as it has been done previously in the literature \citep{Quinn,Spencer2017,Spencer2018}. These values have been obtained by fitting the period distribution of stars between the types F7 and G9 in the solar neighborhood \citep{Duquennoy1991}. 
    In Sect.~\ref{sect:Period-binaryfraction}, we will use other realistic values for $\overline{logP}$ and $\sigma_{logP}$, to see whether some are preferred by the data. For $\sigma_{logP}$, we will use values between 1.7 and 2.9. For the mean of the distribution, we will use as lower limit a value of $\overline{\log P\left[\text{d}\right]}=3.5$, obtained from the analysis of K-type stars and M-dwarf stars in binary systems by \cite{DistP35} (corresponds to separations of 3 AU for 0.8M$_{\odot}$ RGB  stars). As an upper limit, we  adopted $\overline{\log P\left[\text{d}\right]}=5.8$, which was determined by \citetalias{Spencer2018}, they fit the period distribution of the evolved binaries population synthesis for dwarf irregular galaxies made by \cite{DistP58}. 
    
The period distribution will be limited at the lower and upper ends by the minimum and maximum semi-major axis distances, $a$, expected for the type of targets we are considering, RGB stars, and their galactic environment. Using the third law of Kepler, $P^{2}=\frac{4\pi^{2}}{Gm_{1}(1+q)}a^{3}$; this implies a dependence also on $q$. Typically, the minimum value used for the semi-minor axis is $a_{min} =$ 0.21 AU,  that this is the typical radius of RGB stars; it can be obtained by assuming a mass of 0.8 M$_{\odot}$ and surface gravity of 10 cm s$^{-2}$.  Considering the extreme cases of $q=0.1$ and $q=1$, the minimum period is then $\log\left(P_{min}\left[\text{d}\right]\right)=1.57$ for $q=0.1,$ and $\log\left(P_{min}\left[\text{d}\right]\right)=1.44$ for $q=1$. 
    
However, when considering the Roche lobe overflow (RLOF), the less dense star of the binary system (usually the RGB star) will lose material before both components are within 0.21 AU and the equilibrium of the system will be broken. At this moment, either the envelope of the giant star is lost and its core and the companion keep orbiting around the center of mass, or a supernovae type Ia is produced. Simulations of this type of interaction for a 0.88 M$_{\odot}$ star and a compact companion of 0.6 M$_{\odot}$ (similar conditions of what we typically expect to have) have shown that the typical time at which this happens is only of some tens of years \citep{REichardt2019}. Therefore, we do not expect to observe binary stars in the overflow phase and therefore orbiting closer than the Roche limit. 

In the approximation by \cite{Eggleton1983}, the Roche lobe, $r_{R}$, can be calculated as:
    \begin{equation}
        \frac{r_{R}}{a}=\frac{0.49q^{2/3}}{0.6q^{2/3}+\ln\left(1+q^{1/3}\right)}.
    \end{equation}
    The condition for not having overflow is that $r_{R}>0.21$ AU; this translates into a minimum orbital distance of $a_{min}>1.02$ AU for $q=0.1$ and $a_{min}>0.56$ AU for $q=1$. Those values correspond to minimum periods of $\log\left(P_{min}\left[\text{d}\right]\right)=2.60$ for $q=0.1$ and $\log\left(P_{min}\left[\text{d}\right]\right)=2.07$ for $q=1$. We use the limits corresponding to the RLOF approximation in the main part of our analysis\footnote{Since \citetalias{Spencer2018} and \citetalias{Spencer2017} used $a_{min}>0.21$ AU, we will switch to this assumption when comparing directly to their results.}.

    For the maximum value of the semi-major axis, $a_{\rm max}$, we require that the two components of the binary systems are gravitationally bound at all times. We will consider that the two stars are no longer bound when an interaction with another star is produced, that is to say, when $a$ is larger than the mean free path of a star in the galaxy. The mean free path can be obtained as $\left(\pi\ \sigma_{los}\ t\ \lambda \right)^{-1/2}$, where $\sigma_{los}$ is the galaxy's LoS velocity dispersion, $t$ is the average age of the stars, and $\lambda$ is the volume number density of stars. In order to obtain the volume number density of stars, we assume the luminosity density listed in Table \ref{tab:Paramet}, together with the luminosity of the primary being $L/\text{L}_{\odot}=\left(M/\text{M}_{\odot}\right)^{4}$ (relation that works well in our range of masses \cite{Neill1987}), and the fact that with $m_1$ = 0.8 M$_{\odot}$ and with the $q$ distributions examined, each star has an average mass of around 0.4 M$_{\odot}$.  Once we have the mean free path, that is, $a_{max}$ (listed in Table \ref{tab:Paramet}, together with the parameters used to compute it), we can use again the third law of Kepler to compute the value of $P_{max}$, this time it will be different for each galaxy and will depend on q again. The period distributions are shown in the upper right panel of Fig.~\ref{fig:Distribuciones}, and  their lower and upper limits summarized in Table~\ref{tab:Limits}. 
     \begin{table}
        \centering
        \caption{Limits for the period distribution}
        \label{tab:Limits}
        \begin{tabular}{cccccccccccccccc}
        \hline
          &&& & & &  $q$ = 0.1  & & & & & & $q$ = 1 &&& \\
        \end{tabular}
        \begin{tabular}{lcccc} 
                \hline
                  & $\log\left(P_{max}\right)$  & $\log\left(P_{min}\right)$ & $\log\left(P_{max}\right)$ & $\log\left(P_{min}\right)$  \\
                  P units & [days]& [days]& [days]& [days]\\
                \hline
                Sculptor & 6.73& & 6.61 & \\
                Draco & 6.83& & 6.71 &   \\
                Ursa Minor & 6.88& & 6.75  &  \\
                Leo II & 6.51& 2.60 & 6.38 &  2.07  \\
                Carina & 7.21& & 7.08 &   \\
                Fornax & 6.60& & 6.47 &    \\
                Sextans & 6.51& & 6.38 &   \\
                \hline
        \end{tabular}
 \tablefoot{Two extreme cases of $q =0.1$ and $q=1$ are shown. Col.~1: Galaxy name. Col.~2 and 3: Maximum and minimum period for the case of $q=0.1$. Col.~4 and 5: Maximum and minimum period for the case of $q=1$. The minimum periods are for the RLOF approximation.}
\end{table}

     As mentioned previously, the period of binary systems with our target stars as a primary ($e$) is unlikely to be shorter than  $\log\left(P_{min}\left[\text{d}\right]\right)=2.7$, which corresponds to 500 days. For periods longer than this value, the eccentricity distribution is expected to be rather uniform  \citep{Duquennoy1991, Raghavan2010}. However, in our sampling of a uniform distribution, we will take into account that certain combinations of $P$ and $q$ can make the stars collide for very elliptical orbits. This upper limit in the allowed eccentricity is, as a function of the semi-minor axis, $e_{max}=1-(a_{min}/a)$. Therefore, the order we have followed to sample the $e$ distribution (shown in the lower left panel of  Fig.~\ref{fig:Distribuciones}) is to simulate random values of $q$ according to the assumed  distribution of this parameter, then for $P,$ and, finally, for $e$, taking the various dependencies into account.

      The inclination ($i$) describes the angle between the normal to the plane of the orbit and the LoS to the observer. After some trigonometry, one finds that the distribution of $i$ is $f(i)=sin\left(i\right)$ (see middle left panel of  Fig.~\ref{fig:Distribuciones}), with a face-on orbit having $i=0$ rad and an to edge-on orbit $i=\frac{\pi}{2}$ rad.

     The argument of the periastron ($\omega$) is the angle between the ascending line of nodes of the elliptical trajectory of the star and the periastron. Since this orientation is random, the distribution of this parameter is a uniform distribution. We can see the  distribution in the lower panel of Fig.~\ref{fig:Distribuciones}.

    The true anomaly ($\theta$) is the angle between the direction of periastron and the position of the star, from the main focus of the ellipse. For a star with multiple observations, we need to take into account that this parameter changes with time. In order to sample $\theta$, we made use of an auxiliary parameter called mean anomaly ($M$), that is to say, the fraction of an elliptical orbit's period that has elapsed in a time, $t$. So, $M=M_{0}+kt$, where $M_{0}$ is the initial mean anomaly and k is a constant, due to the second law of Kepler, as a star that has an elliptical orbit sweeps out equal areas in equal times. Thus, the parameter k can be computed by using the third law of Kepler $k=\frac{2\pi}{P}$. We sample $M_{0}$ with a uniform distribution. Now, we can relate the mean anomaly of a star with its eccentric anomaly E with the equation $M=E-e\cdot\sin\left(E\right)$, where $e$ is the eccentricity. We solve the equation numerically for $E$ using the function fsolve of the python package SciPy. Then, the true anomaly can finally be obtained with the relation $\cos\left(\theta\right)=\frac{\cos\left(E\right)-e}{1-e\cdot\cos\left(E\right)}$.
All the distributions above have been sampled by implementing a  Monte Carlo rejection method. 

\subsection{Maximum a posteriori method}\label{sec:MAP}
Now, for each $f$, we are going to use the 10,000 simulations to build a likelihood function that describes how well the observational data can be described in terms of the properties of the binary stars population. Then, we evaluate this function using our observational data to obtain the posterior distribution function (PDF) of a given binary fraction. 

Recalling Bayes theorem, we have that:
    \begin{equation}
        P\left(f\left|D,M\right.\right)=\frac{P\left(D\left|f,M\right.\right)P\left(f\left|M\right.\right)}{P\left(D\left|M\right.\right)}.
    \end{equation}
In the above, $P\left(f\left|M\right.\right)$ is the prior probability of a binary fraction f using a model, $M$, which we assume is a uniform distribution between 0 and 1, since we do not have any prior knowledge on the binary fraction.  
$P\left(D\left|f,M\right.\right)$ is the likelihood function, it represents the probability that our observational data is well described by a fraction $f$ using a theoretical model, $M$; this is the function that we have to build using the parameter $\beta$ and the 10,000 simulated mocks for every fraction, $f$.
The next term, $P\left(D\left|M\right.\right)$, is a normalization factor that we will choose so that the integral of the PDF $\left(P\left(f\left|D,M\right.\right)\right)$ is 1. Therefore, since $P\left(f\left|M\right.\right)$ is uniform, the term $P\left(f\left|D,M\right.\right)$ that represents the probability that a fraction, $f,$ fit well our observational data $D$ for a theoretical model, $M$, is directly proportional to the likelihood $P\left(D\left|f,M\right.\right)$. Then, we only need the likelihood function to obtain the PDF on $f$. 

\begin{figure}
        \includegraphics[width=\linewidth]{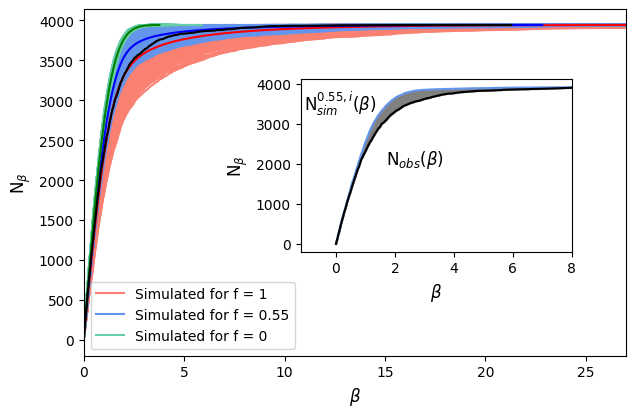}
        \caption{Cumulative distribution function of $\beta$ for Sculptor. The black lines show the CDF obtained for Sculptor from the FLAMES dataset; the CDFs from 750 examples of mock datasets for a binary fraction of 0, 0.55 and 1 are shown in green, blue, and salmon respectively. As there is some overlap between the cases of f = 0.55 and f = 1, the median cumulative curve is highlighted in each case with brighter colors and a thicker line-width. Inset: Example for just one simulation, where in gray we highlight the area between the CDF of the simulated case and that for the Sculptor observations, that is what we use when computing the likelihood function.}
        \label{fig:betaacum}
    \end{figure}

To build our likelihood function, we will be comparing the cumulative distribution function (CDF) of $\beta$ produced by the observations and by the simulated datasets (see Fig. \ref{fig:betaacum}), which we refer to as $N_{obs}(\beta \leq \mathcal{B}$) and $N_{sim}(\beta \leq \mathcal{B}$), respectively: 
    \begin{equation} \label{eq:pdf_f}
        P\left(f\left|D,M\right.\right)\propto\left(\sum_{i=1}^{10000}\int\left|N_{obs}(\beta \leq \mathcal{B})-N_{sim}^{f,i}(\beta\leq \mathcal{B})\right|d\beta\right)^{-\gamma}, 
    \end{equation}
    for all $\mathcal{B}$ positive. Hereafter, for simplicity, we write $N_{sim}(\beta)$ and $N_{obs}(\beta)$ instead of $N_{sim}(\beta \leq \mathcal{B})$ and $N_{obs}(\beta \leq \mathcal{B})$.
    
In Eq.~\ref{eq:pdf_f}, we are considering that the smaller the area between the $N_{obs}(\beta$) and $N_{sim}(\beta{B}$) curves, the better the observations are reproduced by the datasets simulated for a given binary fraction, $f$. The sum is over the 10,000 simulations for each $f$. The integral is performed numerically by summing up the difference between the 3986 sorted values of $\beta$ along the y-axis in Fig. \ref{fig:betaacum}. 

In order to turn the outcome of Eq.~\ref{eq:pdf_f} into an actual probability distribution, we made use of the definite-positive (free) parameter $\gamma$. In practice, we first found the value of $f$ maximizing Eq.~\ref{eq:pdf_f}, with $\gamma = 1$; let's call it $f_{max}$. We then simulated 100 mock datasets with a binary fraction equal to $f_{max}$, using the same simulation strategy explained in Sect.~\ref{sec:Simulation}. Afterwards, we varied the parameter $\gamma$ from 1 to 10 with 0.5 steps (range and precision that we have tested work well), analyzed them as done for the observations, and for each we normalized the PDF. Next, we computed {which value of $\gamma$ is needed so that 68.2\% of the PDFs include $f_{max}$ within 1$\sigma$, 95\% within 2$\sigma$ and 99\% within 3$\sigma$. This way, we make sure that the uncertainties provided with these PDFs are robust.  We note that since $\gamma$ is only an exponent, it will change only the width of the PDF -- not the value of $f_{max}$.

This procedure has to be repeated for every dataset we use (see Sects.~\ref{sec:data_scl} and \ref{sec:data_lit}), since the quality of the data affects the required $\gamma$. Once we have the correct value for $\gamma,$ we only have to compute the $\gamma$ power of our posterior. Finally, we normalized this posterior and we obtained a PDF that peaks at the maximum of the likelihood and that provides well-defined uncertainties. As we see in the next sections, given the lack of this re-normalization step through the parameter $\gamma$, the uncertainties on $f$ can be seriously underestimated.

\subsection{Changes with respect to \citetalias{Spencer2018} methodology} \label{sec:changes}

As discussed, our methodology relies closely on the approach by \citetalias{Spencer2018}, except for three main changes, which we implemented for our analysis of the Sculptor FLAMES dataset, as well as for a re-analysis of literature data for Draco, Ursa Minor, Leo~II, Carina, Sextans, and Fornax. We detail these changes here.

First, \citetalias{Spencer2018} bin the values of $\beta$ and analyzes the distribution of $\beta$ in those bins individually by fitting them with Poisson distributions; in our case, we are using the individual values of $\beta$ and analyzing their CDF. Second, once \citetalias{Spencer2018} computes their PDF, the width of the distribution is not corrected and, therefore, the uncertainties quoted result into formal uncertainties. We note however that the authors provide results from mock simulations (see their Fig. 7), from which it can be appreciated that the formal precision on $f$ has a significantly smaller value than what was suggested by the analysis of mock datasets, even though they do not discuss this implication. In our case, our PDF is corrected so that the quoted 1-,2-, and 3-$\sigma$ uncertainties encompass the 68\%, 95\%, and 99\% of the mock simulations, respectively. Third, \citetalias{Spencer2018} uses a constant value of the minimum orbital distance, fixing it to the average radius of a RGB star: 0.21 AU. In our case, we have also considered the RLOF, so that our minimum orbital distance is the one that makes the Roche-lobe radius larger than 0.21 AU, so that there is no overflow between the components of the binary system.

\section{Results} \label{results}

In Sect.~\ref{sec:scl}, we analyze the results from the application of our method to the Sculptor data presented in Sect.~\ref{sec:data_scl} and test the method on mock datasets, while in Sect.~\ref{sec:lit}, we revise our current understanding on the binary fraction of MW classical dSphs, based on literature datasets.

\subsection{Results from the FLAMES/GIRAFFE Sculptor data} \label{sec:scl}

Figure~\ref{ResultSculptorYO} shows the PDF from the new VLT/FLAMES dataset for Sculptor, both for the case of $a_{min} = 0.21$ AU and when accounting for RLOF. We can see that the median $f$ value returned by the PDF in these two cases is considerably different, even though the sizeable uncertainties bring them in agreement within; essentially, 1$\sigma$ (f = $0.34^{+0.18}_{-0.16}$ in the first case and f = $0.55^{+0.17}_{-0.19}$ in the second case).

As for distinguishing what hypothesis leads to a better representation of the data between $a_{min} = 0.21$ AU and when accounting for RLOF, we can gain information by looking in more detail into the comparison of $N_{obs}(\beta$) with the CDF of $\beta$ produced by the most probable model. Figure~\ref{acumamincomp} shows the comparison to the mean CDF produced by 10000 simulations of the most likely model for each case (f$=$0.55 for the RLOF a and f$=$0.34 for 0.21 AU), separated in different ranges of $\beta$ for ease of visualization.
We can see that for low values of $\beta$s (left panel), the distributions produced under the two assumptions for $a_{min}$ are very similar, where this is because those $\beta$s are mainly due to stars that do not belong to a binary system or that have velocities with large uncertainties. 
In the middle panel, we can see that now both distributions tend to the observational one, with the RLOF case displaying a slightly larger $\beta$ than the 0.21~AU case at a given value on the y-axis, until around \textbf{$\beta=4$}. This is the range where the long period binaries are acting. 
For $\beta,$ from \textbf{4} to 5, we see a change in the behavior, with the red curve (the case for $a_{min}=0.21$ AU), starting to show larger $\beta$s, with respect to the RLOF case at a given N$_{\beta}$. This is the regime where the short period binaries come into play. Finally, in the right panel, showing the regime of larger $\beta$s, the binaries with short periods are dominating, that is why we see a significant difference between the red and blue curves. Among the two, the curve produced with the RLOF approximation stays closer to the observational data than the curve obtained under the assumption of $a_{min} = 0.21$ AU, which departs significantly from the observed behavior. A $\chi^2$ test using bins of size $=1$ and using the CDF of $\beta$ from the observations as the 'expected' distribution yields a $\chi^2 = 329$ for $a_{min}=0.21$ AU, four times larger than that for the RLOF case, $\chi^2 = 75$.  Based on this comparison, we conclude that the RLOF case is to be preferred with respect to fixing the minimal value of the orbital distance to just the radius of the red giant star. Therefore, from now on, all the results shown will be derived under this assumption\footnote{Unless we said otherwise, as when comparing directly to results in the literature that have assumed $a_{min}=0.21$ AU.}. The FLAMES data return a value of f$=0.55^{+0.17}_{-0.19}$ for the binary fraction of Sculptor. 

\begin{figure}
\begin{center}
        \includegraphics[width=\columnwidth]{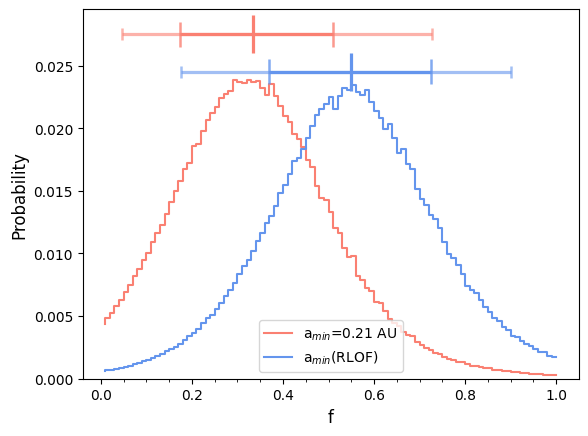}
        \end{center}
        \caption{PDFs obtained for Sculptor, using the VLT/FLAMES dataset. In salmon, we show the result assuming $a_{min}=0.21$ AU and in blue the one taking into account the RLOF, when computing $a_{min}$. The median and confidence levels of 1$\sigma$ and 2$\sigma$ are indicated with horizontal lines. }
        \label{ResultSculptorYO}
    \end{figure}

    \begin{figure*}
    \centering
        \includegraphics[width=1.5\columnwidth]{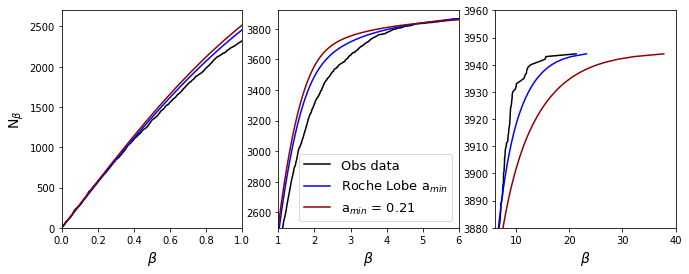}
        \caption{Cumulative distribution function of $\beta$ for Sculptor. In black, we show the observational distribution, in red the one produced by averaging the result of the 10000 simulations for the most likely model using $a_{min}=0.21$ AU and in blue for the RLOF when computing $a_{min}$. The panels show different ranges for the axes in order to better visualize the results.}
        \label{acumamincomp}
    \end{figure*}

    \subsubsection{Testing the method and comparison to S18}

If we concentrate on the results for the $a_{min} = 0.21$ AU case (shown in Figures~\ref{ResultSculptorYO} and in Table~\ref{TablaResultfinal}), which is the assumption adopted by \citetalias{Spencer2018}, it would appear that our approach is less precise than used in the aforementioned study; however, as we will show here, this is just a consequence of our more robust values of the quoted uncertainties on $f$.

To tackle this aspect, we explore what results we would have obtained by applying the methodology by \citetalias{Spencer2018} to our FLAMES dataset for Sculptor, without the modifications we have introduced, and then check with mock datasets the robustness of the uncertainties on $f$.

As a test to check that our implementation of the \citetalias{Spencer2018} method is correct, we run it for Draco and Ursa Minor, using the same datasets as the authors and making the same assumptions for the binary parameter values and distributions. Figure~\ref{resultsSpencer} shows the resulting PDF of the binary fraction, $f$, and Table \ref{tab:ResultsSpencer} summarizes the median of the PDF of the binary fraction together with the 1$\sigma$, 2$\sigma,$ and 3$\sigma$ confidence levels. We can see that our results for Draco and Ursa Minor are in excellent agreement with those obtained by \citetalias{Spencer2018}. The relative differences for $f$ is only on the order of 2\%, which is well within the 1-$\sigma$ error. Those small differences are expected since we changed the size and numbers of the bins used to derive the distribution of $\beta$\footnote{\citetalias{Spencer2018} used 11 different widths between 0.044 and 0.058 and averaged the result, while we use only 1 bin size of 0.025; this is for two reasons: on one hand, it has been shown by \citet{Spencer2018} that the choice of the bin size does not  significantly affect the results; on the other hand, since we have a larger number of $\beta$ in this dataset, with this bin size we make sure that our distributions are well described by a Poisson distribution. 
In addition, in the plots we show, we reduce the noise of the PDFs by averaging pairs of binary fractions.}.

    \begin{figure}
    \centering
        \includegraphics[width=0.8\columnwidth]{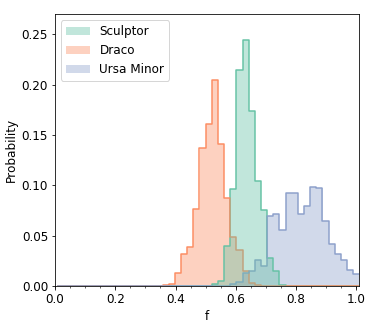}
        \caption{Posterior distribution function for $f$ obtained for Sculptor, Draco and Ursa Minor generated using the methodology by \citetalias{Spencer2018}.}
\label{resultsSpencer}
    \end{figure}

    \begin{table*}
    \centering
    \caption{Binary fraction of Sculptor, Draco, and Ursa Minor and comparaison with literature.}
    \label{tab:ResultsSpencer}
    \begin{tabular}{llllll}
    \hline
    Galaxy     &           & f    & 1$\sigma$          & 2$\sigma$          & 3$\sigma$             \\
    \hline
    Draco      & This work & 0.51 & {[}$-$0.04, +0.05{]} & {[}$-$0.10, +0.10{]} & {[}$-$0.14, +0.14{]}    \\
           & \cite{Spencer2018}   & 0.50 & {[}$-$0.04, +0.05{]} & {[}$-$0.10, +0.10{]} & \multicolumn{1}{c}{-} \\
    \hline
    Ursa Minor & This work & 0.80 & {[}$-$0.09, +0.08{]} & {[}$-$0.17, +0.15{]} & {[}$-$0.15, +0.13{]}    \\
           & \cite{Spencer2018}   & 0.78 & {[}$-$0.09, +0.08{]} & {[}$-$0.15, +0.13{]} & \multicolumn{1}{c}{-} \\
    \hline
    Sculptor   & This work & 0.62 & {[}$-$0.03, +0.04{]} & {[}$-$0.06, +0.08{]} & {[}$-$0.10, +0.11{]}   \\
           & \cite{Spencer2018} & 0.58 & {[}$-$0.17, +0.15{]} & \multicolumn{1}{c}{-} & \multicolumn{1}{c}{-}   \\
           & \cite{Quinn} & 0.59 & {[}$-$0.16, +0.24{]} & \multicolumn{1}{c}{-} & \multicolumn{1}{c}{-}   \\
\hline
\end{tabular}
\tablefoot{Binary fraction for Draco, Ursa Minor and Sculptor using the methodology from \citetalias{Spencer2018}. Col~1 and 2 list the galaxy name and the source of the result, respectively; Col~3. indicates the binary fraction obtained, Cols~4,5, and 6 list the uncertainties for confidence levels of 1$\sigma$, 2$\sigma,$ and 3$\sigma$. It is important to note that the results for Ursa Minor and Draco share the same methodology as well as the same dataset. For Sculptor instead, the S18 methodology is applied to the Sculptor data presented in Sect.~\ref{sec:data_scl}.}
\end{table*}

The application to the FLAMES data returns a much tighter confidence interval for $f$ at a given confidence level (see Table~\ref{tab:ResultsSpencer}), being 0.07 for a confidence level of 68.2\%, to be compared to 0.32 for \citetalias{Spencer2018} and 0.40 for \citet{Quinn} for the same level. This is probably the consequence of various factors. First, the velocity uncertainties are larger in the previous studies, with a median velocity error of  2.1 km s$^{-1}$ versus 1.4 km s$^{-1}$ in our case. Second, even though their sample is larger than ours (190 stars versus 96), our dataset has a larger number of repeated observations and, therefore, of $\beta$s (around 250 from \cite{walker2009} vs 3986). Finally, our maximum baseline is three times larger (12 years versus 4 years).

Nonetheless, we note that the application of the \citetalias{Spencer2018} methodology to mock datasets suggests that the formal uncertainties on $f$ are significantly underestimated. This is  visible in Fig.~\ref{Spen100Scul}, where we show the result for 100 mock datasets built mimicking the characteristics of the FLAMES Sculptor dataset and a binary fraction $f =$ 0.6. We can see that the dispersion in the distribution of the mean of these 100 PDFs is larger than the formal uncertainties that the PDFs themselves provide. Table~\ref{tab:SpencerSims} lists the percentage of models in which the simulated binary fraction $f=0.6$ is within the formal 1 $\sigma$, 2 $\sigma,$ and 3$\sigma$ uncertainties for Sculptor, Draco and Ursa Minor: in all cases, this is much lower than 68\%, 95\%, and 99\% (for example for Sculptor it is 20\%, 28\%, and 42\%). On the other hand, in the same table we list the percentages obtained through the application of our modified methodology and we can see that they are much closer to what expected for 1-, 2-, and 3-$\sigma$ uncertainties.

    \begin{figure}
        \includegraphics[width=\columnwidth]{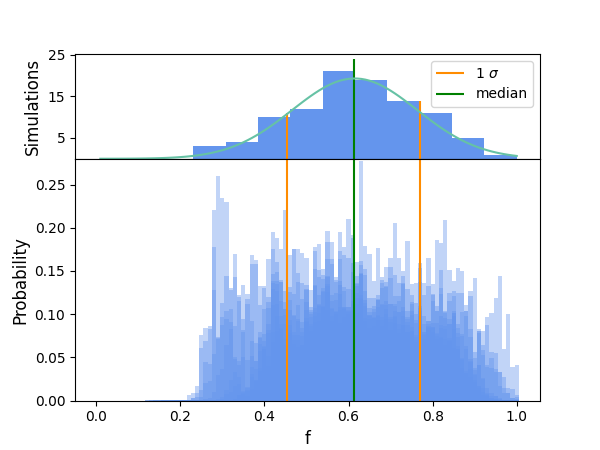}
        \caption{PDFs for 100 mock galaxies simulated reproducing the characteristics of the FLAMES dataset of Sculptor, in terms of number of stars, observations of each star, and velocity uncertainties of each observation, for a binary fraction f$=0.6$. Lower panel: PDFs of the 100 mock galaxies overlaid. Upper panel: Distribution of the maxima of each PDF with a Gaussian fit overlaid. In both panels, the green and orange vertical lines indicate the median of the distribution of maxima and the range that encloses the 68\% of the simulations (1$\sigma$ confidence level).}
        \label{Spen100Scul}
    \end{figure}
    \begin{table}
        \centering
        \caption{Reliable test for the uncertainties.}
        \label{tab:SpencerSims}
        \begin{tabular}{lcr} 
            \hline
                    \citetalias{Spencer2018} method.& \hspace{15pt} & This work \\ 
         \end{tabular}
    \begin{tabular}{lcccccc} 
                \hline
                  & 1$\sigma$ & 2$\sigma$ & 3$\sigma$ & 1$\sigma$ & 2$\sigma$ & 3$\sigma$\\
                \hline
                Sculptor & 20\% & 28\% & 42\%  & 66\% & 92\% & 99\% \\
                Draco & 30\% & 47\% & 59\%  & 68\% & 91\% & 98\%\\
                Ursa Minor & 18\% & 26\% & 38\%  & 68\% & 90\% & 100\%\\
                Leo II & - & - & -  & 66\% & 88\% & 100\%\\
                Fornax & - & - & -  & 67\% & 90\% & 100\%\\
                Carina & - & - & -  & 67\% & 91\% & 100\%\\
                Sextans & - & - & -  & 70\% & 90\% & 98\%\\
                \hline
        \end{tabular}
    \tablefoot{Relative number of simulations that enclose the binary fraction simulated within 1$\sigma$, 2$\sigma,$ and 3$\sigma$ for different galaxies and methods. Col~1. lists the galaxy; Cols~2,3,and 4. indicate the relative number of galaxies for 1$\sigma$, 2$\sigma,$ and 3$\sigma$ relatively using the \citetalias{Spencer2018} methodology; Cols~5,6, and 7. the same for the methodology proposed in this work.}
    \end{table}

     \subsubsection{Period distribution-binary fraction relation} \label{sect:Period-binaryfraction}

\begin{figure*}
\centering
        \includegraphics[width=1.95\columnwidth]{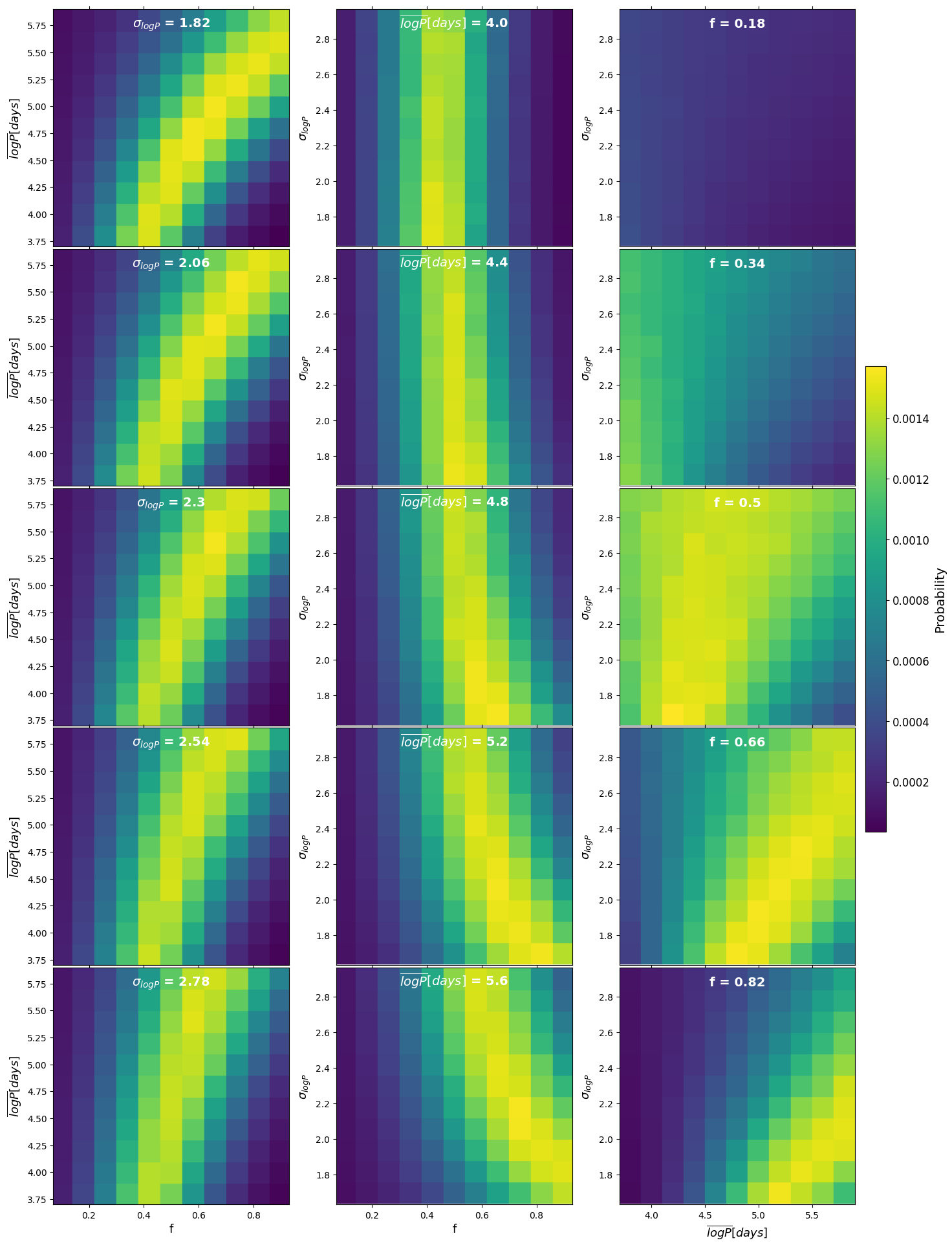}
        \caption{PDFs obtained for the FLAMES Sculptor dataset when simultaneously obtaining two parameters between the binary fraction, $\overline{logP}$ and $\sigma_{logP}$, and keeping the third one fixed to the values in the legend. From left to right, the parameter being kept fixed is the width of the period distribution, the mean of the period distribution and the binary fraction. The probabilities have been normalized so that if we sum a whole column, we obtain 1.}
        \label{degener}
    \end{figure*}
    \begin{figure}
        \includegraphics[width=\columnwidth]{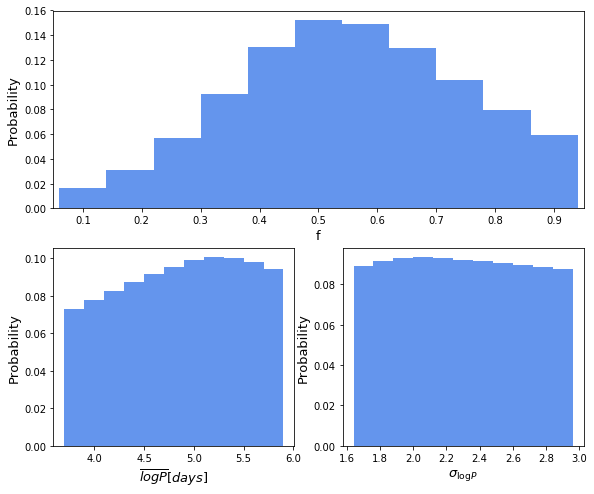}
        \caption{PDFs obtained by summing the colormaps of Fig. \ref{degener}. Here, each bin correspond to the sum of a whole colormap. In the upper panel, we show the PDF for the binary fraction summing all the distribution of periods simulated. In the lower left panel, the one for $\overline{logP}\left[days\right]$ summing all possible binary fractions and $\sigma_{logP}$. In the lower right panel, the one for $\sigma_{logP}$ summing all possible binary fractions and $\overline{logP}\left[days\right]$.}
        \label{colap}
    \end{figure}

Now, we study how the distribution of periods affects the result for the binary fraction. In order to do that, we  use our VLT/FLAMES dataset for Sculptor. We leave the mean and sigma of the period distribution as free parameters, together with the binary fraction. In particular, we explored  11 values of f between 0 and 1, 11 values of $\overline{logP\left[\text{d}\right]}$ between 3.5 and 5.8 and 11 values of $\sigma_{logP}$ between 1.8 and 3, all of them equally spaced. Regarding the value of $\gamma$ used for computing the PDF, we just used $\gamma$=6 since it worked so well in the previous analysis (see confidence intervals for Sculptor in Table  \ref{tab:SpencerSims}). We show the posterior distribution function obtained in Fig. \ref{degener}. We can see that a degeneracy between these parameters exists. In general, we can reproduce the same difference between velocities using lower binary fractions and lower mean periods or larger binary fractions and larger mean periods. This relation is also translated in the dispersion of the period distribution, since a larger dispersion increments the possibilities of producing short period binaries, and those systems matter a lot in terms of contributing to large $\beta$s, so this implies lower binary fractions. This is well in lines with the results obtained by \cite{Quinn}. However, there are three important points that deserve remarks. First, it is not likely to have binary fractions below 0.4 for any period distribution. Second, the different values of $\sigma_{logP}$ produce almost the same shape in the probability distribution. This is because we have constrained the limits of the period distribution properly and then the width of the distribution does not affect the results that much. And third, even though we also have degeneracy in $\overline{logP\left[\text{d}\right]}$ for periods larger than  $\overline{logP\left[\text{d}\right]}=4.8$ , for the lower values we can see that binary fractions between 0.4 and 0.6 are favored for all possible $\sigma_{logP}$. If we look for the most probable case, we obtained a binary fraction of f=0.50, $\overline{logP\left[\text{d}\right]}=4.2$ and $\sigma_{logP}^{2}=1.7$. Even though with the data that we have for Sculptor, this particular case is still the one that better fits the observational CDF of $\beta$, as we checked, we can see that there is still a lot of degeneracy. In particular, for f = 0.50 almost all the possible combinations of periods provided similar results. Therefore, we can not make any strong conclusion about the period distribution if we want to constrain the binary fraction simultaneously. 

We note  that in the work by \cite{Moe2019}, which analyzed the fraction of close binaries in Solar-type stars, found it to have a dependence on metallicity. Rescaling a log-normal distribution to a given observed value of the fraction of stars with companions can account for these results into a period distribution that is not constant with metallicity, with the period distribution shifting towards shorter periods at low metallicity. It could be argued that a distribution with $\overline{logP\left[\text{d}\right]}=4.0$ and $\sigma_{logP}^{2}=2.3$ is better aligned with the period distribution for metal-poor stars in \cite{Moe2019}. In this case, we would find that the binary fraction of Sculptor is $f = 0.34^{+017}_{-0.19}$, but if we look at the CDF of $\beta,$ this leads to a worse fit to the data.  

We have also summed the values of each probability map, to compute the PDF for a certain f, $\overline{logP,}$ or $\sigma_{logP}$ for the combination of the all the other parameters simulated. The results can be seen in Fig.~\ref{colap}. Each bin is the sum of one probability map of the ones shown in Fig.~\ref{degener}, and in the x-axis we have the columns shown in this same Fig.~\ref{degener}. We can see that binary fractions around 0.55 are favored when we combine all possibilities of $\overline{logP}$ and $\sigma_{logP}$. Concerning the period distribution, even though there is noise, we can see that the values of $\overline{logP\left[\text{days}\right]}$ around 5.25 are slightly favored, this could be a hint indicating that we are going in the correct way to constrain it if we achieve better data. About $\sigma_{logP}$ we cannot say anything, it is almost a flat distribution. However, this can indicate that the limits of the period distribution are well selected. With this low-precision result, we cannot properly constrain all the characteristics of the binary star population of Sculptor at once. Nevertheless, as a tentative result, it seems that a similar distribution of parameters of the binary fraction than the one we find in the solar neighborhood for G-dwarfs could work for red giants in DGs, if limited properly. 

\subsection{Application to other MW dSphs} \label{sec:lit}
Analyzing data for Sculptor, Draco, Ursa Minor, Leo~II, Carina, Fornax, and Sextans, \citetalias{Spencer2018} found that the binary star population of MW classical dSph differ in their binary fraction, period distributions or both; for example, if the period distribution does not vary across systems, the binary fractions are spread over a range of values with a width of at least 0.3–0.4. Alternatively, if these systems share the same binary fraction, then they have different period distributions.

 Given the modifications we made to the methodologies used in the literature, we reassess this issue here. Fig.~\ref{resultsextansfornaxcarina} and Table~\ref{TablaResultfinal} display the results of our analysis on literature spectroscopic datasets for Draco, Ursa Minor, Leo~II, Carina, Fornax, and Sextans. We can see that the median does not reproduce well the region of maximum probability for Carina, Fornax, and Ursa Minor. That is why we also show the mode of the distribution and the highest density intervals (HDI). The HDI is, for unimodal distributions as our PDFs, the smallest region around the mode that ensure the correct value is included within a certain confidence level. In our case, the regions are defined as 1$\sigma_{hdi}$, 2$\sigma_{hdi}$ and 3$\sigma_{hdi}$ for confidence levels of 68\%, 95\%, and 99.7\%. In general, the distributions are Gaussian-like, which is good since we expect this behavior for a PDF, with low noise. Even though the results agree with those in the literature \citep{Quinn, Spencer2017,Spencer2018}, within 1$\sigma$, the uncertainties are quite large compared to the results from \citetalias{Spencer2017} and \citetalias{Spencer2018}. However, as it happens in the previous case, now the uncertainties are well derived thanks to the Monte Carlo correction (see the right part of Table \ref{tab:SpencerSims}). We can also see that the new VLT/FLAMES dataset we present for Sculptor is the one that provides the smallest  uncertainties. 
 
 With these new results, there is no evidence for differences in the binary fraction across MW classical dSphs (see also Appendix~\ref{sec:appendix1}). This gives some hope for being able to use dSphs as "templates" for the binary population of UFDs, since these are more challenging systems for which to assemble multi-epoch data for large samples of member stars, on which to compute binary fractions. 
 In order to test possible variations of these parameters in the population of MW DGs, we will have to wait for future datasets (e.g., Sect.~\ref{sec:outlook}). 
    \begin{figure}
    \begin{center}
    \centering
        \includegraphics[width=\columnwidth]{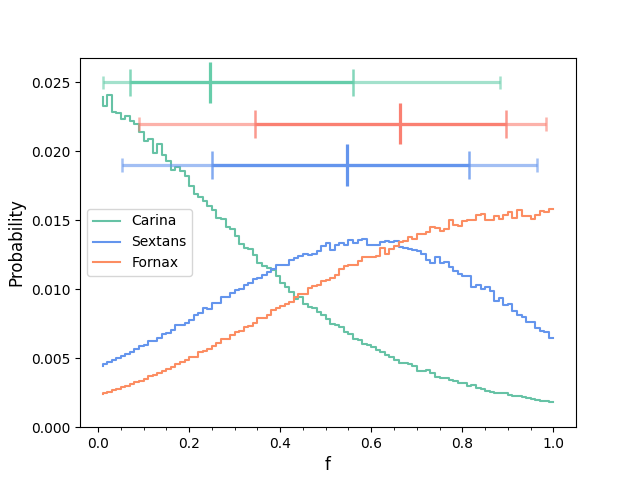}
    \centering
        \includegraphics[width=\columnwidth]{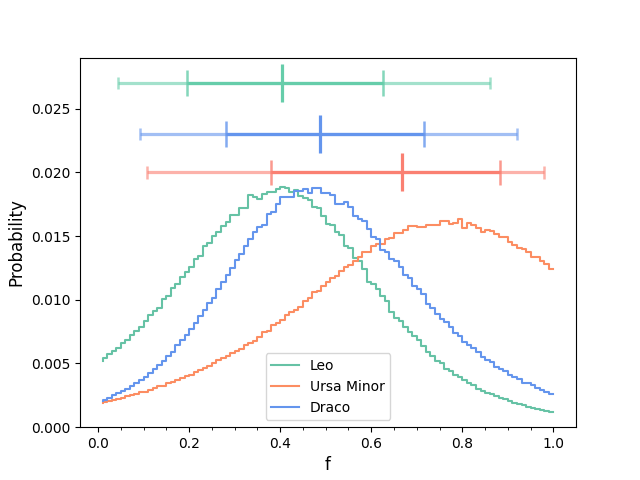}
        \end{center}
        \caption{PDFs on $f$ for the MW dSphs indicated in the legend from our re-analysis of literature datasets, using the new method proposed in this work. 
        The median and 1$\sigma$  and 2$\sigma$ confidence levels are indicated with horizontal lines with ticks.}
        \label{resultsextansfornaxcarina}
    \end{figure}
    
    \begin{table*}
    \centering
    \label{TablaResultfinal}
        \caption{Binary fraction in dwarf galaxies.}
    \begin{tabular}{llllllllll}
    \hline
    Galaxy     & $\gamma$ & f    & 1$\sigma$          & 2$\sigma$              & 3$\sigma$       & $f_{mode}$    & 1$\sigma_{hdi}$          & 2$\sigma_{hdi}$              & 3$\sigma_{hdi}$                        \\
    \hline \\
    Scl   & 7    & 0.55 & {[}0.37, 0.72{]} & {[}0.18, 0.90{]} & {[}0.07, 0.97{]} &
    0.56 &{[}0.38, 0.72{]}&{[}0.20, 0.91{]}&{[}0.07, 1.00{]}\\
    Dra    & 5    & 0.49 & {[}0.28, 0.72{]} & {[}0.09, 0.92{]} & {[}0.01, 0.99{]} &
    0.48 &{[}0.28, 0.70{]}&{[}0.11, 0.93{]}&{[}0.03, 1.00{]}\\
    UMi     & 4.5  & 0.67 & {[}0.38, 0.88{]} & {[}0.11, 0.98{]} & {[}0.01, 0.99{]} &
    0.80 &{[}0.55, 1.00{]}&{[}0.19, 1.00{]}&{[}0.03, 1.00{]}\\
    Leo II    & 5    & 0.40 & {[}0.19, 0.62{]} & {[}0.05, 0.86{]} & {[}0.01, 0.98{]}&
    0.41 &{[}0.19, 0.60{]}&{[}0.10, 0.80{]}&{[}0.01, 0.97{]}\\
    Car   & 4.5  & 0.25 & {[}0.07, 0.56{]} & {[}0.01, 0.88{]} & {[}0, 0.99{]} &
    0.03 &{[}0.00, 0.38{]}&{[}0.00, 0.80{]}&{[}0.00, 0.99{]}\\
    Sxt   & 6    & 0.55 & {[}0.25, 0.81{]} & {[}0.05, 0.96{]} & {[}0.01, 1{]}  &
    0.59 &{[}0.31, 0.86{]}&{[}0.11, 1.00{]}&{[}0.02, 1.00.{]}\\
    Fnx   & 6.5  & 0.66 & {[}0.34, 0.90{]} & {[}0.09, 0.98{]} & {[}0.01, 1{]} &
    1.00 &{[}0.53, 1.00{]}&{[}0.17, 1.00{]}&{[}0.03, 1.00{]}\\  
    \hline
    \end{tabular}
    \tablefoot{Binary fraction parameters obtained by using the new methodology presented in this work (and the RLOF approximation). Col.~1 lists the galaxies. Col.~2 the $\gamma$ parameter (see Eq.~\ref{eq:pdf_f}) that returns realistic confidence intervals; Col.~3 the binary fraction resulting from the median of the PDFs. Cols.~4, 5, and 6 the uncertainty range that correspond to 1$\sigma$, 2$\sigma,$ and 3$\sigma$ confidence levels respectively. Col.~7 the binary fraction resulting from the mode of the PDFs; Cols.~ 8, 9, and 10 the highest density intervals of the PDFs corresponding to 1$\sigma$, 2$\sigma,$ and 3$\sigma$ confidence levels, respectively.}
    
    \end{table*}

\section{Outlook} \label{sec:outlook}
In the previous section, we have seen that currently available spectroscopic datasets for MW dSphs still yield large uncertainties for the binary fraction. The situation worsens even further when allowing for variations in the properties of the binary stellar population; for example, it is well known that a strong degeneracy exists between the binary fraction, as well as the mean and the width of the period distribution \citep{Quinn}.

The next years will see the new generation of wide-field multi-object spectrographs, such as WHT/WEAVE, VISTA/4MOST, and DESI produce exquisite datasets for stars in the Milky Way and its neighbouring galaxies. Here, we explore the improvements that specifically designed datasets could bring to this topic, taking the characteristics and expected performance of the WEAVE spectrograph as template \citep{Jin2023} and Draco as a possible target galaxy. 

We simulated the LoS velocities for seven observations spread over two years, three of them in the first three months and the rest equally spaced in time, in which the same 750 stars are targetted\footnote{The multi-object mode of WEAVE allows for an average 950 fibers per plate; here we conservatively assume that 100 fibers will be devoted to sky and guide stars, and that 100 might remain unallocated.}. To each of these stars we assign an uncertainty on the LoS velocity that depends on the star G-magnitude, as estimated from the analysis of simulated WEAVE spectra in the low resolution mode during part of one of the operational rehearsals \citep{Jin2023}.  This is, errors between 0.4 km s$^{-1}$ for a star with G = 16.5 mag and 2.2 km s$^{-1}$ for one with G = 20 mag. As targets, we choose the 750 brightest stars in Draco, selected from the list of members in \cite{battaglia2022}. All of them have a probability membership above 0.9.
    
 We then simulate 100 mock datasets with a hypothetical binary fraction, $f=0.6$, and then repeat the same analysis outlined in Sect.~\ref{sec:methodology}. The resulting PDFs are shown in Fig. \ref{resultdracoexp}. We obtain that, with a $\gamma=4.5$, 69\% of the models predict the correct binary fraction within 1$\sigma$, 90\% within 2$\sigma,$ and 100\% within 3$\sigma$. For an example PDF, we obtain a binary fraction $f=0.59$ and 1$\sigma$, 2$\sigma,$ and 3$\sigma$ confidence levels of [0.55,0.64], [0.47,0.71], and [0.43,0.77].  With seven exposures over only two years, the 1$\sigma$ uncertainties are now less than half what the current data return. Should the binary fraction of MW classical dSphs cover a range of values with width 0.3-0.4, such differences will be detectable. 

These expected WEAVE datasets will produce much more precise $\beta$ distributions. This could help us to constrain the period distribution of binaries in dwarf galaxies, using the same comparaision we did in Sect.~\ref{sec:scl} (see Fig.~\ref{acumamincomp}). This would be usefull on the degeneracy issue between the binary fraction and the period distribution. It is important to notice that we are far more sensitive to the lower part of the period distribution, so this is the regime that right now one can aim to constrain.

    \begin{figure*}
    \centering
        \includegraphics[width=1.5\columnwidth]{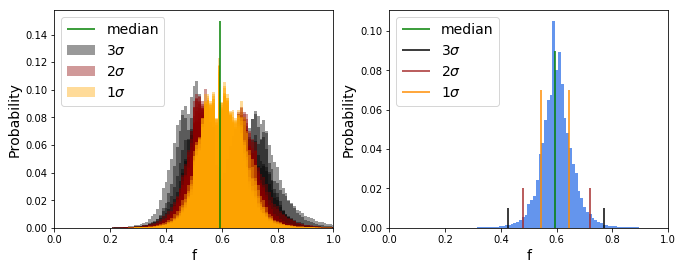}
        \caption{PDFs for simulated WEAVE datasets of Draco. Left panel: PDFs for the 100 simulated WEAVE datasets of Draco (assumed binary fraction f = 0.6) obtained using the new methodology implemented in this work. In orange, we show the cases where the correct binary fraction is obtained within 1$\sigma$ (69\%); in brown within 2$\sigma$ (21\%, up to 90\%) and in black those that predict it within 3 $\sigma$ (10\%, up to a 100\%). The vertical green line indicates the median of the maxima distribution. Right panel: One single PDF of the ones shown in the left panel, this is what we expect to obtain when analyzing the future observational data. Vertical lines indicate the median and intervals of confidence of 1$\sigma$, 2$\sigma,$ and 3$\sigma$, as indicated in the labels.}
        \label{resultdracoexp}
    \end{figure*}

\section{Conclusions} \label{sec:conclusions}

Recently, it has been shown that the binary fraction, $f,$ of MW classical dSphs differ from each other and in some cases deviates from that in the solar neighborhood. Besides the intrinsic interest in understanding the properties of the populations of binary stars in low density and low metallicity environments of MW dwarf galaxies, tackling their binary fraction is also important for determinations of the dark matter content (and, potentially, distribution) of systems at the low-mass end of the galaxy mass function. In fact, LG dwarf galaxies display the largest dynamical mass-to-light ratios and these inferences rely heavily on the $\sigma_{los}$ of the stellar component, which is (more often than not) obtained from single epoch measurements. 

Whilst it is unlikely that the observed values of $\sigma_{los}$ are significantly inflated by unresolved binary stars for dwarf galaxies, with intrinsic $\sigma_{los} > $4 km/s, this might be a serious concern for systems of likely much lower intrinsic, $\sigma_{los}$, as UFDs, with the level of inflation being strongly dependent on $f$. 

We found that some of the uncertainties on $f$ in the most recent works focusing on such determinations for MW dSphs were significantly underestimated. Therefore, in this article, we propose some modifications to such methodology to amend these issues, and test the performance on mock datasets. 

We applied the method to a novel VLT/FLAMES spectroscopic sample for the Sculptor dSphs, that includes repeated observations of the same 96 red giant branch stars with a time baseline of 12 years, with the same instrument, grating (and fiber set-up for the majority of the repeated observations). We also re-analyzed the multi-epoch datasets used before by \citet{Quinn,Spencer2018}. The results are the following:

    Among the most widely assumed hypotheses for the minimum orbital distance between the components of the binary systems, nam equal to the typical radius of a M=0.8 M$_{\odot}$ red giant star ($a_{min}=0.21$ AU) or the larger cut-off as determined by the Roche lobe overflow, the VLT/FLAMES data for Sculptor are best reproduced by the case of the RLOF approximation.

    The analysis of the VLT/FLAMES dataset for the Sculptor dSph returns a binary fraction of $0.55^{+0.17}_{-0.19}$ (1$\sigma$). A result that is in agreement with those from the literature \citep{Quinn,Spencer2018}. Moreover, we have tested our methodology using mock data reproducing the VLT/FLAMES dataset. For this binary fraction the code recovers the correct result 66\%, 92\% and 99\% of the times within 1$\sigma$, 2$\sigma,$ and 3$\sigma,$ respectively. Therefore the uncertainties we provide are well-estimated.

If we reproduce  the methodology by  \citetalias{Spencer2018} precisely, and apply it to the VLT/FLAMES dataset, we obtain a binary fraction for Sculptor $f=0.62_{-0.03}^{+0.04}$ (1$\sigma$). This result is also in very good agreement with those in the literature \citep{Quinn,Spencer2018}, but with our formal uncertainties improved by a factor of 4-5. This is most likely due to the homogeneity of our sample, the larger number of repeated observations and the longer time baseline.  However, in this case it only recover the correct binary fraction 20\%, 28\%, and 42\% of the times within 1$\sigma$, 2$\sigma,$ and 3$\sigma,$ respectively. Therefore, we concluded that these uncertainties are underestimated.
Even though the VLT/FLAMES dataset for Sculptor is the best one available in the literature for the purpose of studying the binary stellar population of a dwarf galaxy, it is still not sufficient for breaking the degeneracy between the binary fraction and the mean and width of the period distribution.
    Our re-analysis of literature spectroscopic datasets for LeoII, Draco, Ursa Minor, Carina, Sextans, and Fornax imply that there is no evidence at present for varying properties of the binary stellar population in these galaxies or from deviations from that of the solar neighborhood.

   \noindent From the point of view of the methodology, our analysis appears to return robust uncertainties. However, there are some improvements that could still be worth implementing.  For example, in the current implementation, one loses all the information about which velocity variations correspond to which star, while it would be interesting to retain this information, because if a star shows large $\beta$ for all the combinations of observations, they are potentially more informative of velocity variation due to binaries than an isolated one that could be caused by other effects. Moreover, we are studying only absolute velocity variations, however, we could also use the information about the sign of the difference between velocities; this could be especially interesting for obtaining periods of individual stars with future datasets. 

    Even though current datasets are still rather unconstraining, it seems likely that it will be possible to shed light on the universality of the binary fraction of MW dSphs in the near future. We forecast that seven repeated measurements of the same 750 brightest stars obtained over only two years in a classical dSph like Draco and an instrument like WEAVE in low resolution MOS mode should deliver a much improved precision on $f$, with 1-$\sigma$ uncertainties about half of those from current spectroscopic data and an uncertainty of $\pm$ 0.2 for a significance of 3$\sigma$. Beside WEAVE and 4MOST, as well as other future instruments (HRMOS, MSE, and WST...) will be able to provide better multi-epoch datasets with even lower velocity uncertainties. However, we are close to a level of precision  in the velocities where the jitter should be taken into account too. In order to make good use of more precise measurements, the jitter should be well characterized, so it may be included in the models.

    \section*{Acknowledgements}
    J. M. Arroyo acknowledges support from the Agencia Estatal de
Investigación del Ministerio de Ciencia en Innovación (AEI-MICIN) and the European Social Fund (ESF+) under grant PRE2021-100638.
    J. M. Arroyo, G. Battaglia, G. Thomas acknowledge support from the Agencia Estatal de
Investigación del Ministerio de Ciencia en Innovación (AEI-MICIN) and the European Regional Development Fund (ERDF) under grant number AYA2017-89076-P, the AEI under grant number CEX2019-000920-S and the AEI-MICIN under grant number PID2020-118778GB-I00/10.13039/501100011033.
The authors acknowledge the referee for the constructive report which enhanced the quality of the manuscript.
    
%
%

    \bibliographystyle{aa}
    \bibliography{ref.bib}
  

\begin{appendix} 

\section{Common binary fraction across MW dSphs} \label{sec:appendix1}
We will check whether we can reject the hypothesis that MW dSphs share the same value of the binary fraction. 
We define as our null hypothesis that all the dwarfs have a binary fraction of 0.5. Under this assumption, all the PDFs that we see in Sect.~\ref{results} would be the result of noise in the measurements. We decide to use 0.5 because all the PDFs have this value in the $1\sigma$ interval, and 4 of them in the $1\sigma_{hdi}$. This is more or less what we expect to have for our sample of seven dwarf galaxies, given that this range indicates a 68.2\% confidence level. 

In order to test our null hipothesis, we will follow the next steps. (1) First, we randomly extracted one value of the binary fraction for each of the dSphs according to their PDFs. Then we subtracted the value that we want to test (f = 0.5) to each of the extracted values.
(2) Next, we carried out a bootstrapping for the seven values generated, exchanging positions 10000 times. Finally, we computed the mean of the value for each of the 10000 cases. We iterated steps one and two 1000 times to sample the whole PDFs.
The distribution of values generated after step three is a Gaussian distribution centered in 0.02 (which indicates that 0.52 is the exact mean of the distributions). The range with a 95\% confidence level around the mean is [-0.23, 0.27]. This means that we can reject our null hypothesis for the values f < 0.27 and f > 0.77 with a 95\% confidence, but not for any value inside that range. However, it is important to point out that this test is biased since our PDFs only allow binary fractions between 0 and 1. Therefore even if we only had flat PDFs the value of 0.5 would still be slightly favored.

This is why we also carried out a more qualitative analysis. We define the metric:

\begin{equation}
    \centering
    D_f = \sum_{i=1}^7(f - f_{dwarf}^i),
\end{equation}
where $f$ is a certain binary fraction that we want to test and $f_{dwarf}^i$ is the mode of the PDF for the {\it i} galaxy. We used this mode because this is the worst case scenario, as the value differs more than the median from 
0.5. Therefore, $D_f$ indicates how close our results are to a single value of f. We will test the binary fraction f = 0.52. The value of  $D_{0.52}$  in our observational case is 1.53. Then, we simulated measurements of a constant binary fraction (f = 0.5) 7 times with the precision given by the 1$\sigma_{hdi}$ uncertainties. So, we can compare if the result we obtain is what we expect if all the dwarfs have a common binary fraction. If we simulate 10000 cases we obtain a Gaussian centered in 1.24 and with a width of 0.35 for the expected values of $D_{0.5}$. This means, that in the case that all the dwarfs have a binary fraction of 0.5, we obtain the expected results given our precision in the measurements, even within 1$\sigma$. Therefore, we cannot neglect the null hypothesis with this method either.
\end{appendix}

\end{document}